\documentclass[]{rsoslite}

\makeatletter
\renewcommand{\p@subsection}{\arabic{section}. }
\makeatother

\usepackage{longtable}  

\usepackage{url}
\usepackage{upgreek}
\usepackage[caption=false]{subfig}
\usepackage{mathrsfs}
\usepackage{xspace}

\newcommand{\Slat}{\ensuremath{\lambda}}     
\newcommand{\Sspt}{\ensuremath{\varsigma}}           

\newcommand{\Sadjmat}{\mathbf{S}}
\newcommand{\Sdistmat}{\mathbf{D}}
\newcommand{\Sprogmat}{\mathbf{P}}

\newcommand{\Sgran}{\ensuremath{\tau}}

\newcommand{\SrecipW}{\bar{\rho}}  
\newcommand{\SrecipT}{\bar{r}}     
\newcommand{\Scofm}{y_\mu}    
\newcommand{\Srofg}{y_\mathrm{gyr}}    

\newcommand{\SrofgAvg}{\bar{y}_\mathrm{gyr}}    

\newcommand{\SinComp}{\mathscr{C}}  

\newcommand{\SeffTime}{\mathcal{E}^\Slat}             
\newcommand{\SeffDist}{\mathcal{E}^\Sspt}             

\newcommand{\Sreach}{K}  

\newcommand{\SrobTime}{R^\Slat}   
\newcommand{\SrobDist}{R^\Sspt}   

\newcommand{\Starr}[1]{  {{t}^{\prime}_{\mathrm{arr}{#1}}}  }  
\newcommand{\Sdist}{{dist}}  

\newcommand{\DXunderground}{London Metro}
\newcommand{\DXparis}{Paris Metro}
\newcommand{\DXnewyork}{New York Metro}
\newcommand{\DXcelegans}{C. Elegans}
\newcommand{\DXflights}{US Flights}
\newcommand{\DXstudentlife}{StudentLife}

\newcommand{\DSunderground}{\textsc{London}\xspace}   
\newcommand{\DSparis}{\textsc{Paris}\xspace}
\newcommand{\DSnewyork}{\textsc{New York}\xspace}
\newcommand{\DSflights}{\textsc{US Flights}\xspace}

\newcommand{\Dunderground}{\textsc{\DXunderground}\xspace}
\newcommand{\Dparis}{\textsc{\DXparis}\xspace}
\newcommand{\Dnewyork}{\textsc{\DXnewyork}\xspace}
\newcommand{\Dcelegans}{\textsc{\DXcelegans}\xspace}
\newcommand{\Dflights}{\textsc{\DXflights}\xspace}
\newcommand{\Dstudentlife}{\textsc{\DXstudentlife}\xspace}

\newcommand{\node}[1]{\emph{#1}}

\newcommand{\gfxscale}{0.87}  %
\newcolumntype{C}[1]{>{\centering\let\newline\\\arraybackslash\hspace{0pt}}m{#1}}
\usepackage{multirow}
\newcommand{\rom}[1]{\uppercase\expandafter{\romannumeral #1\relax}}  

\usepackage{fix-cm}

\begin{document}

\title{Spatio-temporal networks: reachability, centrality, and robustness}

\author{
Matthew J. Williams$^{1,2}$ and Mirco Musolesi$^{2,1}$}

\address{%
$^{1}$School of Computer Science, University of Birmingham, Edgbaston B15 2TT, UK\\
$^{2}$Department of Geography, University College London, London WC1E 6BT, UK}



\corres{Matthew J. Williams\\
\email{m.j.williams@cs.bham.ac.uk}}

\begin{abstract}


\noindent

Recent advances in spatial and temporal networks have enabled researchers to more-accurately describe many real-world systems such as urban transport networks. In this paper, we study the response of real-world spatio-temporal networks to random error and systematic attack, taking a unified view of their spatial and temporal performance. We propose a model of spatio-temporal paths in time-varying spatially embedded networks which captures the property that, as in many real-world systems, interaction between nodes is non-instantaneous and governed by the space in which they are embedded. Through numerical experiments on three real-world urban transport systems, we study the effect of node failure on a network's topological, temporal, and spatial structure. We also demonstrate the broader applicability of this framework to three other classes of network. To identify weaknesses specific to the behaviour of a spatio-temporal system, we introduce centrality measures that evaluate the importance of a node as a structural bridge and its role in supporting spatio-temporally efficient flows through the network. This exposes the complex nature of fragility in a spatio-temporal system, showing that there is a variety of failure modes when a network is subject to systematic attacks.


\end{abstract}

\maketitle

\newcommand{\mcghosts}[1]{}
\mcghosts{\includegraphics{openaccesslogo_bw.pdf}}
\mcghosts{\includegraphics{OPENSCIENCE.pdf}}
\mcghosts{\includegraphics{RS_crossmark_logo.pdf}}
\mcghosts{\includegraphics{RSOS_Pubs_Logo_Line_CMYK.pdf}}



\section{\label{sec:introduction}Introduction}

Network science provides many powerful methods to study a great variety of systems in society, nature, and technology. Modelling complex systems in terms of their network structure allows researchers to understand, predict, and optimise their real-world behaviour~\cite{newman:2010:networks:,newman:2006:structure}. Detailed data describing the interactions and relationships in real-world systems have become increasingly available, emerging from domains as diverse as transport~\cite{gallotti:2015:multilayer}, biology~\cite{bota:2015:architecture,chen:2006:wiring}, infrastructure~\cite{yook:2002:modeling,sole:2008:robustness}, and sociology~\cite{eagle:2009:inferring}.
The ability of network analysis to capture relationships and dependencies between components makes it an essential tool for studying system resilience~\cite{albert:2000:error,cohen:2000:resilience,de_domenico:2014:navigability,holme:2002:attack,callaway:2000:network}.
It is known that local disruptions can have a significant impact on the global behaviour of a system~\cite{albert:2000:error,albert:2004:structural}.
The study of a network's vulnerabilities and its response to malfunction helps engineers design robust systems~\cite{albert:2004:structural,berezin:2015:localized,buldyrev:2010:catastrophic,berche:2009:resilience,sole:2008:robustness,dallasta:2006:vulnerability} and scientists understand complex phenomena such as neurodysfunction~\cite{fornito:2015:connectomics,stam:2014:modern}, economic and financial risk~\cite{helbing:2013:globally,thurner:2013:debtrank-transparency:}, and disease spreading~\cite{schneider:2011:suppressing}.

Methods for robustness analysis generally assume that the system is represented by a network composed of static links, focusing on the topological properties of a network subject to disruption. Depending on the system and research question, a static representation may also incorporate weighted and directed edges, allowing richer dynamics to be modelled. In many systems, however, edges are not continuously active~\cite{holme:2012:temporal} and the quantities their weights represent may vary with time. Furthermore, these time-varying systems may also be spatially embedded, and thus the ability for nodes to interact is governed by the space in which they operate as well as their network connectivity. 

Urban transport systems have a long history in network analysis~\cite{gallotti:2014:anatomy,de_domenico:2014:navigability,angeloudis:2006:large,lee:2008:statistical,derrible:2010:complexity,roth:2012:long-time}, and are an exemplar of a system that is temporal~\cite{holme:2012:temporal}, spatial~\cite{barthelemy:2011:spatial}, and multilayer~\cite{kivela:2014:multilayer}. Examples of recent studies of transport systems that account for time-varying and multilayer properties, either jointly or separately, can be found in Refs. ~\cite{gallotti:2014:anatomy,de_domenico:2014:navigability,pan:2011:path,roth:2012:long-time}. Furthermore, the resilience of such systems is of particular interest, with insights from static network analysis enabling better-engineering of, for example, rail and metro systems~\cite{derrible:2010:complexity,dorbritz:2009:stability}.
Aside from transport systems, there are many classes of network that are spatial, temporal, or both.
Engineering spatial network resilience in fixed communication networks has particularly received attention over recent years, including new methods for identifying critical geographic regions~\cite{trajanovski:2015:finding,neumayer:2015:geographic,neumayer:2011:assessing} and understanding spatial damage to fibre optic networks~\cite{agarwal:2013:resilience}.
Communication networks and cellular nervous systems are just few examples of systems that can be represented by means of networks embedded in space and time. Relying only on a static space-agnostic aggregation of such networks over-simplifies the rich and complex relationships in the real systems they represent.

The consequences of ignoring the temporal and spatial constraints on networks have been highlighted in recent surveys of spatial~\cite{barthelemy:2011:spatial}, temporal~\cite{holme:2012:temporal}, and multilayer~\cite{kivela:2014:multilayer,boccaletti:2014:structure} networks.
Advances in temporal networks have provided researchers with a valuable framework that can be used to understand the temporal structure of systems, and empirical measurements have demonstrated that static representations often over-estimate the true connectivity of real-world temporal systems~\cite{konschake:2013:robustness,trajanovski:2012:error,karsai:2011:small,kivela:2012:multiscale}. Furthermore, while static methods investigate purely topological measures of system vulnerability, such as giant component size and network diameter, new methods allow researchers to additionally explore temporal efficiency~\cite{scellato:2013:evaluating} and expose differences in the temporal function of the network when subject to random error versus systematic attack~\cite{sur:2015:attack,trajanovski:2012:error}.

However, temporal network models sometimes make the strong assumption that interactions are instantaneous. For many types of spatially embedded system, this assumption ignores the influence that space has in constraining the structure of the network. As an example, transit between stations in a public transport system naturally incurs a time delay while a passenger travels, and the specific delay depends on the speed of the service currently operating and distance between stations. 
In one of the earliest works on temporal networks, Kempe et al.~\cite{kempe:2000:connectivity} propose a schedule-based analysis of transport systems as a potential application of temporal network analysis; more-recent application of time-respecting paths~\cite{kempe:2000:connectivity,holme:2005:network} to transport networks can be found in Pan and Saram\"{a}ki~\cite{pan:2011:path}, and Gallotti and Barthelemy~\cite{gallotti:2014:anatomy}.
Critical nodes in such a system are those that not only bridge two clusters in the network, but also act as a conduit for \textit{rapid} flow between physically distant areas of the system.

To study robustness in spatially embedded temporal systems we apply a general framework that is able to capture instantaneous and non-instantaneous types of interaction. We formulate a model of spatio-temporal systems in which the interactions and relationships between components are constrained by the space and time in which they are embedded (Sec.~\ref{sec:framework}). 
The analysis of a network's spatio-temporal structure (defined in Sec.~\ref{sec:framework}) provides the foundation to measure its topological, temporal, and spatial function. These measures are presented in Sec.~\ref{sec:attacks}, along with systematic attack strategies designed to expose different weaknesses in the network. 
In order to validate our framework and show its utility and flexibility, in this paper we explore the behaviour and resilience of six empirical examples of spatio-temporal systems (described in Sec.~\ref{sec:datasets}); specifically, three urban transit systems, a national air travel network, a biological neural network, and a mobile phone communication network.
Our empirical analysis in Sec.~\ref{sec:analysis} highlights a crucial distinction when examining the robustness of spatio-temporal systems: we must consider the impact of disruption to a network's temporal and spatial efficiency, in addition to its topological structure. Furthermore, critical nodes play different roles in terms of their topological, temporal, and spatial utility, and therefore systematic attack strategies can differ in the damage they cause to the network.




\section{\label{sec:framework}Connectivity in spatio-temporal networks}

We start by introducing the notation and framework with which we define connectivity in a spatio-temporal network. 
Our model follows a fundamental property of spatial networks: the space in which the system is embedded acts as a constraint on the structure of the network~\cite{barthelemy:2011:spatial}. In a spatio-temporal setting, we can represent this constraint as the speed with which one node can interact with another. This is a natural abstraction for many real-world processes that have previously been modelled as purely static or temporal systems; for example, passenger transit in public transport systems, neurotransmission in biological neural networks (e.g., Ref.~\cite{grindrod:2010:evolving}), shipping in multimodal freight networks (e.g., Ref.~\cite{kaluza:2010:complex}), movements in trade networks (e.g., Ref.~\cite{bajardi:2011:dynamical}), and signal delay in telecommunications networks (such as the Internet). 
Our approach mixes structure (i.e., topology), space, and time, and allows for an exploration of the influence of each dimension on the processes occurring in these networks.

We summarise the relevant notation introduced throughout Sec.~\ref{sec:framework} to Sec.~\ref{sec:attacks} in Table~\ref{notation}.

\begin{table*}
\caption{\label{notation}%
Summary of notation used in this paper.}
\centering
%
\begin{tabular}{l|l}
\hline
Symbol & Description
\\
\hline

$\Sgran$            & Window width \\
$G^{[t]}$           & Graph for snapshot at time $t$ \\
$\Sadjmat^{[t]}$    & Matrix of physical propagation speeds between nodes for snapshot at time $t$ \\
$\Sdistmat^{[t]}$   & Matrix of physical distances between nodes for snapshot at time $t$ \\
$\Sreach^{[t]}$     & Reachability set (dependent on implicit origin node $v_0$ and start time $t_1$) \\
$\Sprogmat^{[t]}$   & Progress matrix (dependent on implicit origin node $v_0$ and start time $t_1$) \\
$\Slat$             & Temporal efficiency \\
$\Sspt$             & Spatial efficiency \\
$S$                 & Size of largest strongly connected component as fraction of all nodes \\
$\SrobTime(f)$      & Temporal robustness at deactivation rate $f$ \\
$\SrobDist(f)$      & Spatial robustness at deactivation rate $f$ \\
\hline
\end{tabular}%
\end{table*}

\subsection{Time-varying spatially embedded systems}

A time-varying network is conventionally represented as a time-ordered sequence of graphs~\cite{kempe:2000:connectivity,nicosia:2013:graph,holme:2014:analyzing,holme:2013:temporal,holme:2012:temporal}, with each graph corresponding to a snapshot of the network during a particular time window. The time intervals are commonly finite and equally sized, and we refer to the interval duration as the \emph{temporal granularity} $\Sgran$. We assume an overall observation duration that consists of $T$ timesteps, starting with an initial time $t_1$.

More formally, let $V$ denote the set of all nodes in the system under study, and let $N=|V|$ be the number of nodes in the system.
We consider a temporal graph consisting of $T$ discrete non-overlapping windows, represented by the time-ordered sequence of directed graphs $G^{[t_1]}, \ldots, G^{[t_T]}$.
A particular graph $G^{[t]} = (V,E^{[t]})$ captures the topological state of the spatio-temporal system during the interval $[t, t + \Sgran)$.
It is assumed that nodes are present throughout the lifetime of the system; that is, each graph has the same node set $V$.
For each graph $G^{[t]}$, where $t = t_1,\ldots,t_T$, there exists a counterpart weight matrix $\Sadjmat^{[t]} \in \mathbb{R}^{N \times N}$ that represents the weighted directed edges between nodes during the interval $[t, t + \Sgran)$. The weight of the edge from node $v$ to node $w$ at time $t$ corresponds to the element $\Sadjmat_{vw}^{[t]}$. In our model, $\Sadjmat_{vw}^{[t]}$ is a non-negative scalar representing the speed of physical propagation from $v$ to $w$ in the corresponding time interval.
For example, in a public transport system, $v$ and $w$ are typically transit stations and $\Sadjmat_{vw}^{[t]}$ is the average speed at which passengers are conveyed from $v$ to $w$ given the service operating during $[t, t + \Sgran)$. In the case that there is no connection from node $v$ to $w$ at time $t$, we have $\Sadjmat_{vw}^{[t]}=0$. 

In a spatio-temporal system, the constituent entities naturally occupy a location in space, and this location may vary over time. More formally, we assume nodes are embedded in a $k$-dimensional metric space with physical distance function $g$. The embedding is dependent on the system; for example, geographic networks are commonly represented in a spherical (or ellipsoidal) coordinate system (where $g$ is geodesic distance over the surface of the Earth) or projected on to a two-dimensional Euclidian space.
We write $l^{[t]}_v$ to denote the physical position\footnote{In practice, a non-stationary node may occupy many locations during this interval, and the method for choosing a single representative location $l^{[t]}_v$ depends on the system being studied.} of node $v$ during the interval $[t, t + \Sgran)$, noting that $l^{[t]}_v \in \mathbb{R}^k$.
For convenience, we collect the time-varying pairwise distances between nodes in a physical distance matrix $\Sdistmat^{[t]}$, where $\Sdistmat^{[t]}_{vw} = g( l^{[t]}_v, \, l^{[t]}_w )$ for each $v,w \in V$.
We make the simplifying assumption that multiple nodes in the network cannot occupy the same location. In real-world systems, nodes represent physical entities, and thus cannot occupy the same space.

\subsection{\label{sec:propagation_proc}Model description}

To apply classical network-theoretic concepts in a spatio-temporal system, we use the notion of space-time constrained propagation between nodes. Paths through the spatio-temporal network are defined by this process, which then form the basis for higher-order network measures such as connected components and network distance. These paths obey the time-varying conditions of the system along their route.
This form of connectivity follows the same spreading process that is common in defining temporal paths~\cite{nicosia:2013:graph,holme:2014:analyzing,holme:2013:temporal}, with the modification that propagation from one node to another is constrained by the speed of transmission between nodes and their physical distance. This framework's explicit encoding of time-varying speeds serves as an alternative to event-based representations of similar networks (e.g., as applied to an air transport network in Ref.~\cite{pan:2011:path}).

We consider propagation as a discrete-time process starting at an origin node $v_0$ and the initial time $t_1$, and progressing over each subsequent timestep $t_2, t_3, \ldots, t_T$. 
In the following description of the propagation process we treat the origin node $v_0$ and start time $t_1$ as implicit parameters.
Modelling non-instantaneous propagation in a time-varying system necessarily involves capturing partial propagation between nodes, as well as the nodes that have been reached along a path.
The process is represented by two time-evolving structures: a \emph{reachability set} $\Sreach^{[t]}$ and an $N$-by-$N$ \emph{progress matrix} $\Sprogmat^{[t]}$. The reachability set $\Sreach^{[t]}$ consists of nodes that have been reached from origin node $v_0$ by the end of timestep $t$. 
The progress matrix $\Sprogmat^{[t]}$ represents the distance accumulated during direct propagation between nodes; more specifically, the element $\Sprogmat^{[t]}_{vw}$ gives the \emph{progress} from node $v$ to node $w$ at timestep $t$ measured as the distance accumulated during propagation from $v$ to $w$. A progress element represents the state of partial propagation between the two nodes at the end of a timestep.
In this process, progress from node $v$ to $w$ is able to accumulate while the two nodes are continuously connected by an edge. The amount by which progress is incremented in each timestep is governed by the propagation speed between the two nodes. Given a sufficient period of continuous connectivity between $v$ and $w$, progress can accumulate to the point where it exceeds the physical distance between the two nodes, thereby representing complete propagation from $v$ to $w$.

To give an example, consider the case of a national rail network where the elements of the network are cities and edges are transport links.
Let us consider transportation with vehicles travelling between two cities $v$ and $w$ at an average speed of 200 km/hour. If the distance between the cities is 150 km, after a 15-minute time interval (indicated by $\Sgran=$ 15 minutes) the value $\Sprogmat^{[t]}_{vw}$ will be equal to 50 km. Thus, if the average speed remains constant, then three timesteps must elapse before $w$ is reachable from $v$.

\begin{figure*}
\centering
\includegraphics[width=\textwidth]{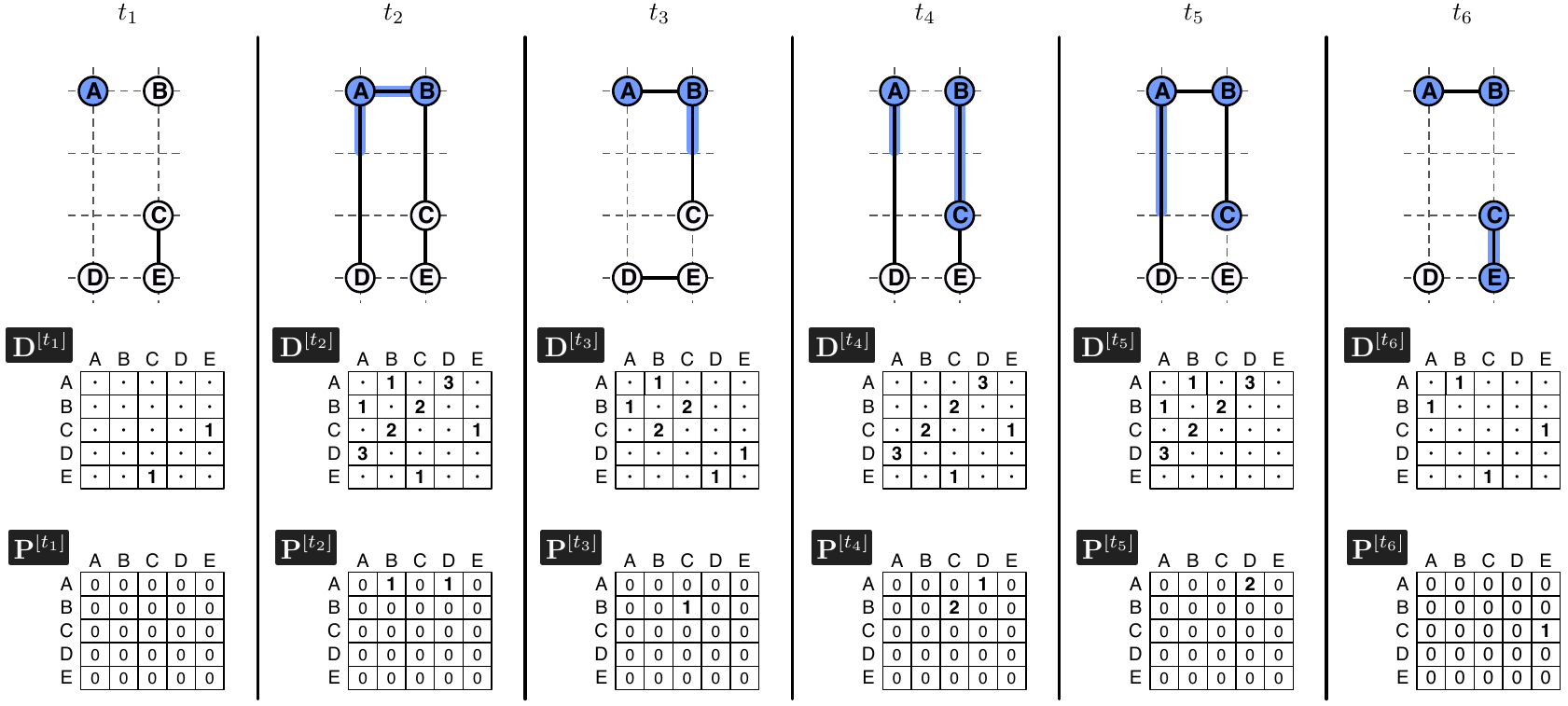}%
\caption{\label{fig:stnet_schematic}Example of the constrained propagation process in a spatio-temporal system. Propagation begins with origin node \node{A} at timestep $t_1$. Nodes are positioned at integer locations on a Euclidian plane. Temporal granularity $\Sgran=1$ second. All edges have a propagation speed of 1m/s. Membership of a node in the reachability set $\Sreach^{[t]}$ at the end of timestep $t$ is represented by shading. The corresponding distance matrix $\Sdistmat^{[t]}$ and progress matrix $\Sprogmat^{[t]}$ are shown below each network snapshot. Progress is also visually represented by shading along an edge. For clarity, we omit from $\Sdistmat^{[t]}$ distances between pairs of nodes that are not connected in $G^{[t]}$.
}
\end{figure*}

Let us now consider the more-complex example in Fig.~\ref{fig:stnet_schematic}. Here we can see that source node \node{A} reaches \node{B} and \node{C} in at most three subsequent timesteps. Specifically, full propagation from \node{A} to \node{B} is able to occur in one timestep, arriving at timestep $t_2$. Then, in the case of propagation from \node{B} to \node{C}, the distance between the nodes is larger, and thus more time is required to complete the interaction. We therefore see that there is an intermediate stage in timestep $t_3$ where there is only partial propagation from \node{B} to \node{C}. This completes in timestep $t_4$. The example also demonstrates two failed propagation attempts between \node{A} and \node{D}. In $t_3$ propagation is set to zero due to the absence of an edge from \node{A} to \node{D}. The reappearance of the edge in timestep $t_4$ allows propagation to restart, but subsequently fails again at $t_6$.

We now formalise the propagation process.
The initial state of the process is such that origin node $v_0$ is the only reachable node and no progress has been made; that is, at the initial timestep $t_1$ we have $\Sreach^{[t_1]}=\{v_0\}$ and $\Sprogmat^{[t_1]}$ is all zero.
When progressing from timestep $t_{i-1}$ to $t_i$, where $i \geq 2$, further propagation from a reachable node $v$ to an unreachable node $w$ depends on whether an edge from $v$ to $w$ exists in timestep $t_i$. 
If these conditions hold, the distance by which we can increment progress from $v$ to $w$ is the \emph{propagation increment} $\Sgran \cdot \Sadjmat^{[t_i]}_{vw}$, where $\Sadjmat^{[t_i]}_{vw}$ is propagation speed and $\Sgran$ is the window duration.
We note that the propagation increment may exceed the \emph{remaining distance} required to propagate from $v$ to $w$, which we denote by $q^{[t_i]}_{vw}$. The quantity $\Sgran \cdot \Sadjmat^{[t_i]}_{vw}$ therefore represents the maximum amount that may be added to the progress of $v$ to $w$, while the actual increment is given by
\begin{equation}
\min(\Sgran \cdot \Sadjmat^{[t_i]}_{vw},\, q^{[t_i]}_{vw})\,.
\end{equation}
The remaining distance $q^{[t_i]}_{vw}$ is obtained from the current physical separation distance $\Sdistmat^{[t_i]}_{vw}$ and previous progress $\Sprogmat^{[t_{i-1}]}_{vw}$ as
\begin{equation}
q^{[t_i]}_{vw} = \Sdistmat^{[t_i]}_{vw} - \Sprogmat^{[j-1]}_{vw}\,,
\end{equation}
or set to 0 if this quantity becomes negative due to a change in the pair's positions between the two timesteps.

Finally, we derive the progress matrix update rule. The value  $\Sprogmat^{[t_i]}_{vw}$ in timestep $t_i$, where $i \geq 2$, is
\begin{equation}\label{eqn:progress_update}
\Sprogmat^{[t_i]}_{vw}
  = \Sprogmat^{[t_{i-1}]}_{vw} + \min(\Sgran \cdot \Sadjmat^{[t_i]}_{vw},\, q^{[t_i]}_{vw})
\end{equation}
if $v \in \Sreach^{[t_{i-1}]}$, $w \not\in \Sreach^{[t_{i-1}]}$, and edge $(v,w)$ exists in $G^{[t_i]}$. In the other case that the system as no edge from $v$ to $w$ at timestep $t_i$, progress is reset and so $\Sprogmat^{[t_i]}_{vw}$ is set to 0.
Propagation from $v$ to $w$ is considered successful in timestep $t_i$ if the cumulative progress $\Sprogmat^{[t_i]}_{vw}$ between the nodes exceeds or equals their physical distance $\Sdistmat^{[t_i]}_{vw}$. In such a case, $w$ is regarded as reachable from source node $v_0$ at timestep $t_i$, and $w$ is included in the set $\Sreach^{[t_i]}$. More formally, the reachability set at timestep $t_i$ is expressed as
\begin{equation}
\label{eqn:reachability_update}
\Sreach^{[t_i]} = \Sreach^{[t_{i-1}]} 
  \, \cup \,
  \{\,
    w \,\, | \,
    \, \exists \, v
    \; \text{s.t.} \;
    \Sprogmat^{[t_i]}_{vw} \, \geq \, \Sdistmat^{[t_i]}_{vw}
  \,\} \;.
\end{equation}

In the preceding equation, $\Sprogmat^{[t_i]}_{vw} \, \geq \, \Sdistmat^{[t_i]}_{vw}$ represents the case that sufficient time has elapsed for propagation to complete, expressed in terms of the cumulative propagation distance $\Sprogmat^{[t_i]}_{vw}$ with respect to the current separation $\Sdistmat^{[t_i]}_{vw}$ between the two nodes. Hence, the process models the effect of nodes that may be non-stationary in their spatial embedding. This means that, for example, propagation is accelerated when two nodes are moving towards one another.

To summarise, the spatio-temporal constrained propagation process is entirely described by the time-evolution of three matrices: a physical distance matrix $\Sdistmat^{[t]}$ of spatial distances between nodes; a propagation speed matrix $\Sadjmat^{[t]}$, which specifies the speed of transmission between pairs of connected nodes; and a progress matrix $\Sprogmat^{[t]}$, which tracks incremental propagation between nodes and evolves according to Equation~\ref{eqn:progress_update}.
Node-to-node propagation is successful only after sufficient time has elapsed, and this duration is a result of the timestep width $\Sgran$, the time-dependent distance between nodes, and the time-dependent speed between nodes.

As with discrete-time temporal network representations (e.g., \cite{nicosia:2012:components,fenu:2015:block,grindrod:2010:evolving,alsayed:2015:betweenness,tang:2010:small-world}), the temporal granularity $\Sgran$ represents a necessary trade-off between abstraction and fidelity. In our propagation process, finer granularities minimise the amount of excess in the propagation increment on successful propagation (i.e., $\Sprogmat^{[t_i]}_{vw} - \Sdistmat^{[t_i]}_{vw}$). Too-coarse a granularity leads to under-estimation of the true propagation duration, and over-estimation of the propagation distance.
An upper bound for the quality of the choice of the parameter $\Sgran$ can be obtained by considering the physical distances and speeds of the underlying system. In particular, we find the minimum non-zero value of the product $\Sdistmat_{vw}^{[t]} \cdot \Sadjmat_{vw}^{[t]}$ over all $t=t_1, \ldots, t_T$ and $v,w \in V$. 
For the given choice of $\tau$, this quantity represents the minimum direct propagation duration that exists in any of the system's snapshots. If the minimum direct propagation time is greater than $\tau$, it means that the representation is guaranteed to under-sample the temporal dynamics of the network. 

We have so far considered propagation through the system from a single origin node $v_0$, starting at timestep $t_1$, and proceeding through timesteps $t = t_2, \, t_3, \, \ldots, \, t_T$. For network analysis, this allows us to construct spatio-temporal paths originating at $v_0$. The main features from the propagation process we consider are the time at which a node $w$ was reached from $v_0$, and the physical distance travelled through intermediate propagation along the route from $v_0$ to $w$. By repeating this process from each node in $V$ we construct system-wide network connectivity in the time interval $[t_1, t_T + \Sgran)$, thereby yielding spatio-temporal paths starting at time $t_1$.

\subsection{Spatio-temporal paths and distance measures}

In general, a \textit{spatio-temporal path} from node $v_0$ may visit multiple distinct vertices before reaching its destination node. A spatio-temporal path consisting of $n \geq 0$ hops, starting with origin node $v_0$ at timestep $t_1$, is described as the sequence of $n+1$ pairs
\begin{equation}\label{eqn:st-path}
	\left\langle\,
	(v_0, \, t_1),
	\,
	(v_1, \, \Starr{_1}),
	\,
	(v_2, \, \Starr{_2}),
	\,
	\\
	\ldots,
	\,
	(v_n, \, \Starr{_n})
	\,\right\rangle,
\end{equation}
where $v_j$ denotes the $j$th node visited on the path and $\Starr{_j}$ denotes the time at which the path reached node $v_j$.
Spatio-temporal paths are readily constructed by tracing the propagation process described in the previous section.

In addition to the timestep $\Starr{_j}$ in which the path arrives at a node $v_j$, the propagation process also captures the physical distance traversed in reaching $v_j$ from preceding node $v_{j-1}$, given by the progress matrix element
\begin{equation}
\Sprogmat^{[\Starr{_j}]}_{v_{j-1}v_{j}} \, .
\end{equation}
This quantity is useful for studying the distance travelled along the route.
We also note that any such path constructed from the propagation process also obeys the temporal-ordering condition 
\begin{equation}
t_1 <  \Starr{_1} < \ldots < \Starr{_n} \; .
\end{equation}
Furthermore, as is the case with temporal networks, reachability in spatio-temporal networks is non-transitive and non-symmetric.

To illustrate an example, in Fig.~\ref{fig:stnet_schematic} we can see that node \node{A} is able to reach nodes \node{B}, \node{C}, and \node{E}. From this we can construct the spatio-temporal routes taken to reach each node from \node{A}. For example, to reach \node{E} we can trace the path from \node{A} through nodes \node{B} and \node{C} and identify the total time taken as five seconds and overall physical distance travelled along the path as four metres.

Throughout the rest of the paper our notation will omit the observation window $[t_1,\, t_{T} + \Sgran)$ in which the spatio-temporal paths obtained from a spatio-temporal system, leaving it as an implicit parameter. We stress, however, that the following measures we introduce are defined on the paths present in such an interval.

Temporal paths are characterised by two notions of length~\cite{nicosia:2013:graph,holme:2012:temporal}: the \emph{topological length}, which is the number of hops along the path\footnote{Specifically, topological length is the number of hops along the (temporally) shortest path, which may not necessarily be a path with the minimum number of hops between the two nodes.}, and \emph{latency}, which is the time elapsed from source to destination. In reference to the previous description of a spatio-temporal path (Equation~\ref{eqn:st-path}), topological length corresponds to $n$ and latency is given by $\Starr{_n} - t_1$.
In the context of a spatio-temporal path, an additional relevant feature is the \emph{spatial length} of a path from $v_0$ to $v_n$, given by
\begin{equation}
\sum_{j=1}^{n} \Sprogmat^{[\Starr{_j}]}_{v_{j-1}v_{j}}
\,.
\end{equation}
These quantities allow us to explore connectivity in the network from topological, temporal, and spatial perspectives.

To study the robustness of the system we focus on the shortest paths between nodes, as these represent optimal routes within the system. Indeed, the typical length of such paths is assumed to be representative of the overall efficiency of the network.
As commonly defined in temporal networks \cite{nicosia:2013:graph,holme:2012:temporal}, a path from node $v$ to $w$ is a temporally shortest path if it has minimum latency.  We extend this definition to our spatio-temporal setting. Formally, a path from $v$ to $w$ is a \emph{spatio-temporally shortest path} if it is a temporally shortest path and has minimum spatial length.

Although multiple temporally shortest paths may exist from one node to another, it is a subset of these paths that are spatio-temporally shortest, and all such paths share the same latency and spatial length. Under this definition, notions of spatio-temporal distance in the network have a well-defined meaning. Specifically, the \emph{temporal distance}
$\Sdist^\Slat_{vw}$
from $v$ to $w$ is the latency of the spatio-temporal shortest path from $v$ to $w$. Similarly, the \emph{spatial distance}
$\Sdist^\Sspt_{vw}$
from $v$ to $w$ is the spatial length of the spatio-temporal shortest path from $v$ to $w$. In the case that $w$ cannot be reached from $v$ in the given time interval (and thus no path exists between them) both distance quantities are set to $\infty$.

\label{sec:st_paths:compute}
In practice, to compute spatio-temporal shortest paths in a network, we follow an epidemic (i.e., breadth-first) approach, similar to the Brandes method~\cite{brandes:2001:faster} for computing betweenness centrality in weighted static networks. As with the Brandes algorithm, we must enumerate shortest paths through the network; however, due to the non-transitive and non-symmetric nature of temporal and spatio-temporal paths, intermediate shortest paths cannot be re-used in the calculation of broader paths.
Here we construct a breadth-first shortest path tree for each node in the network. The tree is rooted at the given node, and spatio-temporal shortest paths from the root are obtained according to the process defined in Sec.~\ref{sec:propagation_proc}.
In our implementation, computing each tree can require a worst-case time of $\mathcal{O}(N^2)$ where $N$ is the number of nodes.
In our practical experiments (see datasets in Sec.~\ref{sec:datasets}), we parallelise this tree-construction process, and find that it is feasible to extract shortest paths in networks up to a size of 400-450 nodes, though note that runtime also depends on the sparsity of the network.




\section{\label{sec:attacks}Random error and systematic attack}

In order to study the resilience of a network we define different ways in which nodes are selected to fail. 

\emph{Random error} (abbreviated to \emph{Err}) refers to probabilistic models of node failure. In this paper, we consider uniform random failure, where each node has an equal independent failure probability of $f$, yielding an expected number of node failures $f \cdot N$. Random error allows us to measure the typical response of the network to random breakdowns. It is also a useful comparison for systematic attack, as any intelligent attack strategy should be at least as harmful to the network as random error.

Systematic attacks on networks rely on the ability to strategically identify and deactivate nodes that are critical to the function of the system. Although many measures of node centrality are available to define attack strategies~\cite{newman:2010:networks:}, only a subset of these are effective in rapidly disrupting the behaviour of a network~\cite{trajanovski:2012:error,pan:2011:path}. Attack on \emph{spatio-temporal} networks adds further complexity, as the importance of a node can be evaluated in terms of its centrality in the topological, temporal, and/or spatial structure of the system, properties which are not necessarily correlated.

In the rest of this section we introduce relevant centrality measures and formalise the attack strategies we use to test the robustness of a spatio-temporal network. We also define the quantities with which we measure the performance of a network.

\subsection{Spatio-temporal centrality measures}

A variety of formulations of closeness in temporal networks have been developed. Given the non-symmetry of connectivity and latency in temporal networks, we must distinguish between in- and out-closeness. In particular, common definitions include temporal out-closeness~\cite{pan:2011:path} and temporal in-farness~\cite{trajanovski:2012:error,sur:2015:attack}. The latter has been proposed as a basis for a centrality measure in systematic attacks on networks~\cite{trajanovski:2012:error,sur:2015:attack}. Here we present temporal in-closeness as the reciprocal of temporal in-farness. Formally, given the giant temporal in-component $\SinComp_v$ of node $v$, we can consider the average temporal distance of nodes that reach $v$ as
\begin{equation}
\ell_v =
  \frac{1}{|\SinComp_v| - 1}
  \sum_{w\in\SinComp_v;\,w\not=v} \Sdist^\Slat_{wv}
  \, .
\end{equation}
A node with low $\ell_v$ is considered more-central in the network. An unreachable node has no utility in the network, and thus we set $\ell_v=\infty$ if $\SinComp_v$ is singleton. The \emph{temporal in-closeness} of a node $v$ is then given by
\begin{equation}
C^{\mathrm{TC}}_v = \frac{1}{\ell_v} \, .
\end{equation}
Inverting in-farness simply reverses the ranking of nodes, ensuring high values for more-central nodes.

However, while closeness is useful for identifying nodes capable of rapidly reaching the rest of the network (e.g., for fast patching of computer malware~\cite{tang:2012:stop:}), it does not directly measure the reliance of the network on a node as an efficient connector, which is often a critical source of network fragility. 
This notion is better represented by betweenness centrality~\cite{holme:2002:attack}, which we extend to our spatio-temporal network as follows. First, we denote the set of spatio-temporally shortest paths from node $w$ to node $u$ by $\sigma_{wu}$ and let $\sigma_{wu}(v)$ denote the subset of paths in $\sigma_{wu}$ that pass through node $v$. \emph{Path betweenness centrality} is then defined for a node $v$ as
\begin{equation}
C^{\mathrm{PB}}_v = \sum_{w,u \,\in\, V; \, w\not=u} \frac{|\sigma_{wu}(v)|}{|\sigma_{wu}|} \,. 
\end{equation}
Although this quantity represents the importance of a node in supporting connectedness within the network, it does not necessarily reflect a node's potential role as a hub through which information can be quickly transferred. Indeed, efficient temporal conduits such as these are especially important for rapid transmission within a network (as recently highlighted in Ref.~\cite{alsayed:2015:betweenness}), and deactivating these conduits can have significant impact on the global efficiency of the network. We therefore define the \emph{betweenness efficiency centrality} of a node $v$ as
\begin{equation}
C^{\mathrm{BE}}_v = \sum_{(w,u) \, \in \, \beta_{v}} \frac{1}{\Sdist^\Slat_{wu}} \,,
\end{equation}
where $\beta_{v}$ is the set of source-destination pairs for each path in $\sigma_{wu}(v)$. The reciprocal distance $1/\Sdist^\Slat_{wu}$ is useful as a measure of the efficiency of the shortest path $w$ to $u$, and is commonly used in evaluating the network efficiency of static~\cite{latora:2001:efficient,latora:2003:economic} and temporal~\cite{tang:2009:temporal,nicosia:2013:graph,holme:2012:temporal} networks.

\subsection{Attack strategies}
We summarise our five chosen attack strategies as follows.
Temporal closeness attack (TC) deactivates nodes in decreasing order by temporal in-closeness centrality. This follows the order defined in the initial (intact) network, similar to the approach in Ref.~\cite{trajanovski:2012:error}.

\begin{figure}
\centering
\includegraphics[scale=\gfxscale]{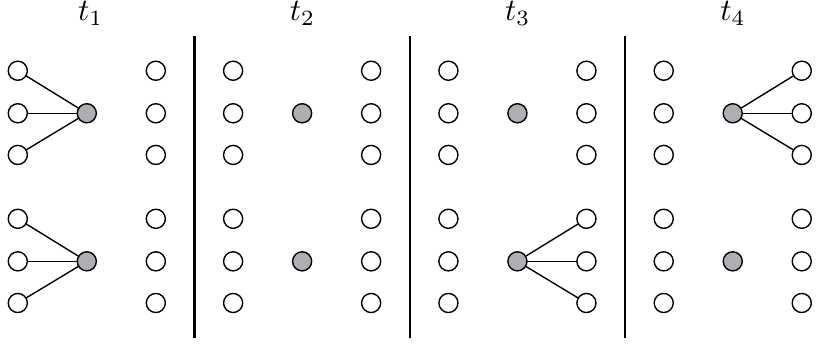}%
\caption{\label{fig:betweenness_demo} Comparison of path betweenness and betweenness efficiency. Although both shaded nodes have the same path betweenness, the bottom node has higher betweenness efficiency. Deactivating the bottom node has a larger impact on the overall temporal performance of the network.}
\end{figure}

Path betweenness attack (PB) and betweenness efficiency attack (BE) exploit path betweenness centrality and betweenness efficiency centrality, respectively. As shown in Ref.~\cite{holme:2002:attack}, the geodesic paths in a network can significantly change after each failure. Thus, in both betweenness-based attacks, the centrality ranking is recalculated after each node failure. 

Fig.~\ref{fig:betweenness_demo} demonstrates the difference in how PB attack and BE attack prioritise their node deactivations. Deactivating either shaded node will have a similar effect on the network's spatio-temporal connectivity; however, the bottom node is supporting paths that have faster propagation flow. Betweenness efficiency attack prioritises nodes which are bottlenecks with high temporal load.

Finally, we include two attacks based on the degree distribution of the intact network: in-degree attack (ID) and out-degree attack (OD). Unlike the other three attack strategies, the ID and OD strategies do not rely on global computation of the spatio-temporal paths in a system. Specifically, we extract in-degrees and out-degrees from the unweighted static aggregate of the original temporal network.

\subsection{Vulnerability measures}

Analysis of robustness depends on the indicators used to quantify network performance. Here we introduce the measures we use to evaluate the topological, temporal, and spatial vulnerability of a spatio-temporal network. While there are many properties relevant to the study of the function of a network, we are careful to select measures that do not confound these three properties.

An important property of real-world networks is the size of the largest strongly connected component, as this represents the extent to which the network is mutually reachable. This is commonly used to study topology in static and temporal contexts~\cite{holme:2002:attack,albert:2000:error,sur:2015:attack}. The notion of a connected component in a spatio-temporal network follows from the temporal definition in Ref.~\cite{nicosia:2012:components}. Formally, a strongly connected component $A$ is a set of nodes where there exists a spatio-temporal path between all pairs of nodes $v,w \in A$. The \emph{giant strong component size} $S$ is the size of the largest strongly connected component as a fraction of the overall number of nodes $N$. The definition of \emph{giant weak component size} follows similarly. However, we focus our robustness analysis on strong component size as this measure is especially relevant in many real-world systems; for example, the resilience of the giant strong component represents the ability of a public transport system to retain mutual navigability when subject to station closures. We note that in this paper we deal with giant components in finite-size real-world systems.

Efficiency~\cite{latora:2001:efficient,latora:2003:economic} gives us a method to explore the overall spatial and temporal performance of a network. For a chosen distance measure (i.e., temporal $\Slat$ or spatial $\Sspt$), efficiency tells us how effective a network is in supporting rapid transmission with minimum distance travelled. Formally, \emph{temporal efficiency} (first defined in Ref.~\cite{tang:2009:temporal}) is the average reciprocal latency over all pairs of nodes in the network, expressed as
\begin{equation}
\SeffTime = \frac{1}{N(N-1)} \sum_{v \not= w} \frac{1}{\Sdist^\Slat_{vw}} \, .
\end{equation}
This quantity is normalised between 0 and 1. In our spatio-temporal framework, $\SeffTime=1$ is achieved if direct propagation between all pairs of nodes occurs in one timestep. On the other hand, $\SeffTime=0$ if propagation does not succeed between any nodes, which can occur if there are no transmission links, or propagations speeds are insufficient to complete direct propagation between nodes.

Our examination of network vulnerability focuses on the system's response to the failure of one or more nodes, which has previously been measured in temporal networks using temporal robustness~\cite{trajanovski:2012:error}.
In this paper we study the effect of complete deactivation of a node; i.e., where the failed node no longer has any connections to or from the rest of the system.
Consider a set of deactivated nodes $D\subseteq V$. We denote the temporal efficiency of a network with deactivated nodes $D$ by $\SeffTime(D)$. The \emph{relative efficiency} is then given by
${\SeffTime(D)}\,/\,{\SeffTime}$.
When evaluating a systematic attack, the choice of nodes $D$ depends on the chosen attack strategy. An effective strategy is one that can cause significant damage with few node deactivations, thus we are particularly interested in the relative efficiency with respect to the fraction $f$ of nodes deactivated, $f=|D|/N$. Formally, we refer to this as the network's \emph{temporal robustness} $\SrobTime(f)$ after the fraction $f$ of nodes have been deactivated. In probabilistic failure models, such as uniform random deactivation, $\SrobTime(f)$ represents the expected relative temporal efficiency given failure probability $f$. (In general, measuring robustness through the relative decline in a global network quantity is sometimes referred to as the $R$-value of that quantity~\cite{trajanovski:2013:robustness}.)

While temporal efficiency measures the latency of the network, its spatial counterpart, \emph{spatial efficiency} $\SeffDist$, is calculated over the reciprocal spatial distances between nodes in the network. Node failure will force paths to take alternative routes in the network, typically over longer physical distances, therefore reducing spatial efficiency. We denote a network's \emph{spatial robustness} with respect to a given deactivation rate $f$ by $\SrobDist(f)$.

\subsection{Computation of attack strategies and measures}

In terms of computation, temporal closeness (TC), path betweenness (PB), and betweenness efficiency (BE) attacks all require calculation of spatio-temporal shortest paths in the underlying network (see Section~\ref{sec:st_paths:compute} for more information). Similarly, measurement of efficiency depends on the shortest distances between each pair of node in the network.
Due to the non-transitive and non-symmetric nature of spatio-temporal paths, in practice we must use the affine graph method to compute giant component sizes (see Ref.~\cite{nicosia:2012:components} for details), which is the basis for the giant strong component size vulnerability measure.
The two local-knowledge attack strategies (ID and OD) are simpler to compute, requiring only knowledge of the aggregate degree distribution.




\section{\label{sec:datasets}Empirical spatio-temporal networks}

In this paper we analyse the spatio-temporal robustness of transport, biological, and social systems through six real-world networks.
Here we detail how each network is constructed. Specific choices of temporal granularity, number of snapshots, and observation duration for each network can be found in Table~\ref{tab:datasets_summary}. The table also includes descriptive summaries of each network's properties.
Further detail on the data materials used in the preparation of each network can be found in the Electronic Supplementary Material. This also includes visualisations of the time-evolution of each network.
\begin{table*}
\caption{\label{tab:datasets_summary}%
Summary of the six spatio-temporal systems explored in this paper: timetabled London Underground transits (\Dunderground), Paris Metro transits (\Dparis), New York City Subway transits (\Dnewyork), US domestic flights (\Dflights), the nervous system of C. Elegans (\Dcelegans), and mobile communications among university students (\Dstudentlife). For each network we show the number of nodes $N$, number of directed edges $|E^*|$ in the aggregate static graph, temporal granularity $\Sgran$ (also referred to as window size), number of timesteps $T$, temporal duration, topological temporal correlation $C$ (Ref.~\cite{tang:2010:small-world}), topological reciprocity $\SrecipT$, weight reciprocity $\SrecipW$, average radius of gyration $\SrofgAvg$ over all nodes, and median propagation speed over all edges. See appendices for definitions of reciprocity (App.~\ref{sec:reciprocity}) and radius of gyration (App.~\ref{sec:gyration}) in a spatio-temporal system.}
%
\resizebox{\textwidth}{!}{%
\begin{tabular}{l|cc c c c ccc c c}
\hline
Network                        &
$N$                          &  
$|E^*|$                        &  
\multicolumn{1}{c}{$\Sgran$}   &  
$T$                            &  
\multicolumn{1}{c}{Duration}   &  
\multicolumn{1}{c}{$C$}          &  
\multicolumn{1}{c}{$\SrecipT$}   &  
\multicolumn{1}{c}{$\SrecipW$}   &  
\multicolumn{1}{c}{$\SrofgAvg$}   &  
\multicolumn{1}{c}{Med. speed}     
\\
\hline
\Dunderground  &
  270  &  628  &  2 mins  &  1440  &  2 days  &
  0.526  &  
  0.77  &  0.71  &
  0.00 m  &  8.39 m/s
  \\
\Dparis  &
  302  &  705  &  2 mins  &  1440  &  2 days  &
  0.444  &  
  0.76  &  0.68  &
  0.00 m  &  6.56 m/s
  \\
\Dnewyork  &
  417  &  1058  &  2 mins  &  1440  &  2 days  &
  0.347  &  
  0.50  &  0.48  &
  0.00 m  &  7.09 m/s
  \\
\Dflights  &
  299  &  3947  &  30 mins  &  480  &  10 days  &
  0.382  &  
  0.42  &  0.38  &  
  0.00 m  &  152.81 m/s
  \\
\Dcelegans  &
  279  &  2990  &  10ms  &  1200  &  12s  &
  1.000  &  
  0.47  &  0.45  &
  0.00 m  &  0.44 mm/s
  \\
\Dstudentlife  &
  22  &  68  &  60 mins  &  1008  &  42 days  &
  0.007  &  
  0.78  &  0.78  &
  18.9 km  &  inst.
  \\
\hline
\end{tabular}%
}
\end{table*}

\textbf{London Underground} (\Dunderground): 
The London Underground is a rapid transit transport system that covers much of Greater London. In our spatio-temporal construction of the Underground system, we model the transits of passenger trains moving between each station along their timetabled routes. Each node is a metro station. We build a fine-grained time-varying representation of station-to-station transit speeds from the vehicles' scheduled journey times and distances. In a given snapshot, we obtain the propagation speed between two consecutive stops by averaging the speeds of trains serving those stations during the corresponding time interval. We select the timetable of February 2015 and set the observation start time to Monday at 00:00.

\textbf{Paris Metro} (\Dparis): 
We obtain a spatio-temporal construction of the Paris Metro using the same approach as for \Dunderground, resulting in a network consisting of 302 stations. We select the timetable of December 2015 and set the observation start time to Monday at 00:00.

\textbf{New York Subway} (\Dnewyork): 
Our spatio-temporal construction of the New York City Subway also follows the same approach as \Dunderground. We select the weekday timetable of December 2015 and an observation start time of Monday 00:00. Multiple stations comprising the same station complex are unified into one node, yielding 417 subway stations active in the observation period.

\textbf{US Domestic Flights} (\Dflights): 
The flights network is constructed from actual take-off and landing times of domestic passenger flights in the United States in the month of February 2014. Each node is a US airport. Our approach to extracting transit speeds is similar to \Dunderground. Time zones are normalised to EST, and we start our observation window at Monday 00:00. Airports are spatially embedded on the WGS-84 ellipsoid and physical distances are calculated using Vincenty's equation.
We note that, unlike the \Dunderground dataset, the flight times are from reported data, rather than \emph{a priori} timetables. Thus, the flight durations reflect environmental conditions that may delay or hasten transits.

\textbf{\DXcelegans} (\Dcelegans): 
\emph{Caenorhabditis elegans} (C. Elegans) is a nematode and the first organism to have its entire cellular nervous system mapped \cite{white:1986:structure}, including its growth from embryogenesis~\cite{varier:2011:neural,nicosia:2013:phase}, spatial configuration~\cite{kaiser:2006:nonoptimal}, and wiring~\cite{chen:2006:wiring,varshney:2011:structural}. In our spatio-temporal construction of the C. Elegans network, we study signalling among the neurons in the adult worm. 
Neuron spatial coordinates were collected in Ref.~\cite{kaiser:2006:nonoptimal} and are given in two-dimensions along the worm's lateral plane.
Missing neuron coordinates were supplemented using the approach described in Ref.~\cite{varier:2011:neural}.
The topological structure of the system is constructed using the synaptic connection data compiled in Ref.~\cite{chen:2006:wiring} and Ref.~\cite{varshney:2011:structural}.
The connectivity data includes the type (chemical or electrical) and density of directed synaptic connections, allowing us to infer transmission delay between neurons.
The task of empirically measuring synaptic latency in C. Elegans is particularly challenging and so far only a small number of synapses have been examined. State-of-the-art optogenetic methods show that synaptic delay over chemical-only synapses is on the order of milliseconds \cite{lindsay:2011:optogenetic}, and specifically in the range 10ms to 30ms (see Sup. Fig. 1 in Ref.~\cite{lindsay:2011:optogenetic}). We encode propagation speeds in our network according to this range of baseline measurements, scaling speed values based on the synapse type (chemical, electrical, or both) and physical distance. Thus, we differentiate between these two type of connection (i.e., electrical and chemical).

\textbf{Dartmouth StudentLife Experiment} (\Dstudentlife):
The StudentLife experiment~\cite{wang:2014:studentlife:} used continuous smartphone monitoring to follow a cohort of students at Dartmouth College over one academic semester. To study the information sharing potential of this network using our spatio-temporal framework we construct the patterns of communication between students via their SMS and phone logs. Each node corresponds to a student participating in the experiment. Communication events between phones correspond to opportunities for instantaneous information transmission between students. Contact between two students in a particular timestamp maps to a link with infinite propagation speed. In particular, calls are represented as bidirectional edges and SMS messages are treated as directed from sender to recipient. Students' mobility patterns are extracted from their GPS and wireless access point geolocation logs. Our observation window covers six weeks from Monday 8 April 2013.

Our framework offers an alternative approximation of passenger transit (i.e., in \DSunderground, \DSparis, \DSnewyork, and \DSflights) to other, space-agnostic temporal models. They capture the speed at which passengers can be conveyed, as well as dependence on time-ordering.
We note that the aggregated directed versions of these two networks have over 99\% reciprocity, whereas their average reciprocity in the spatio-temporal representation is much lower (Table~\ref{tab:datasets_summary}). This highlights the loss of information caused by time aggregation. The effect is especially pronounced in \Dflights, where we see that simultaneous inbound and outbound flights between two airports are rare.
We also note the variation in temporal correlation over the datasets. \Dcelegans is the only network with a static topology, giving a correlation of $1$. 

\begin{figure}
\centering
\includegraphics[scale=\gfxscale]{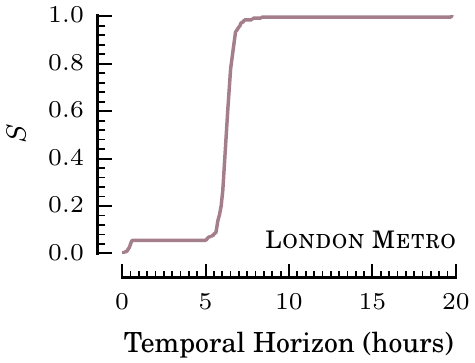}%
\hspace{0.3cm}%
\includegraphics[scale=\gfxscale]{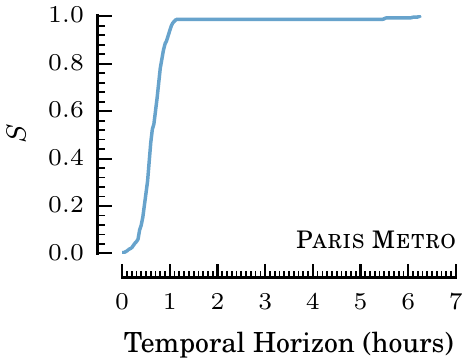}%
\hspace{0.3cm}%
\includegraphics[scale=\gfxscale]{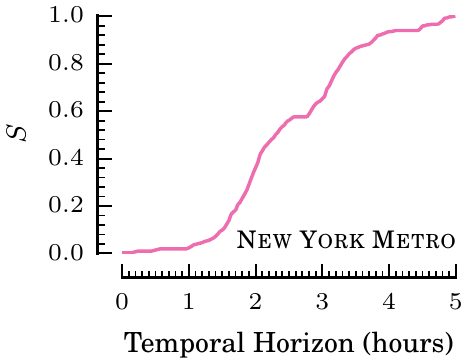}%
\\\vspace{0.25cm}
\includegraphics[scale=\gfxscale]{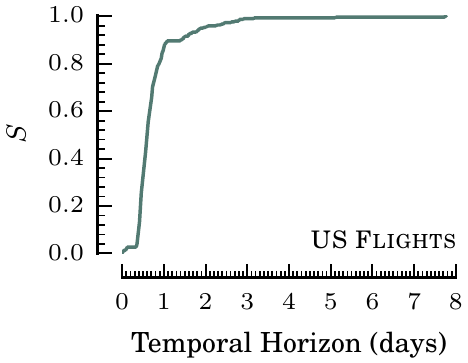}%
\hspace{0.3cm}%
\includegraphics[scale=\gfxscale]{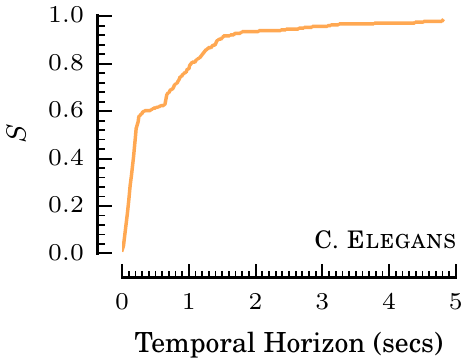}%
\hspace{0.3cm}%
\includegraphics[scale=\gfxscale]{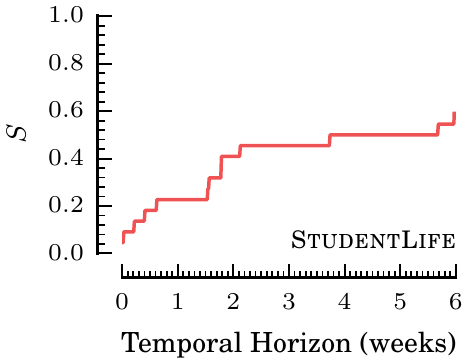}%
\caption{\label{fig:giant_size_temp_horiz} (Colour) Influence of temporal horizon on the giant (strong) component size $S$. The temporal horizon $h$ limits the observation window during which spatio-temporal paths are extracted to $[t_1, t_{1}+h)$.}
\end{figure}

Fig.~\ref{fig:giant_size_temp_horiz} shows the growth of reachability in each spatio-temporal network as the temporal horizon is extended.
In the case of the two transport networks, we see the formation of spatio-temporal paths during normal operating hours, contrasting with little or no growth during early morning. \DSnewyork is unlike the other two metro systems as it operates 24/7, with a slightly reduced night service between midnight and 6am. We also note the effect of different temporal and spatial scales present in each network. The Underground reaches full coverage within a day, whereas neural network communication is on the order of seconds. Although most of the US flight network is mutually reachable within one day, a number of remote off-mainland airports with limited service are responsible for delaying full coverage by up to a week.
Due to the presence of five isolated nodes in the reconstruction of the C. Elegans nervous system we observe that the giant strong component in \Dcelegans does not reach full coverage. This is also observed in aggregate analyses of C. Elegans~\cite{varshney:2011:structural}.




\section{\label{sec:analysis}Results and discussion}

In this section, we use the framework outlined in Sec.~\ref{sec:framework} to compare the response of real-world spatio-temporal networks (described in Sec.~\ref{sec:datasets}) when subject to the node failure models presented in Sec.~\ref{sec:attacks}. Before considering systematic attack strategies, we first explore the response of each network to random error.

\subsection{Tolerance to random error}

\begin{figure*}
\centering
\subfloat[Giant component size]{\label{fig:4sys_rand:giant_strong}%
\includegraphics[scale=\gfxscale]{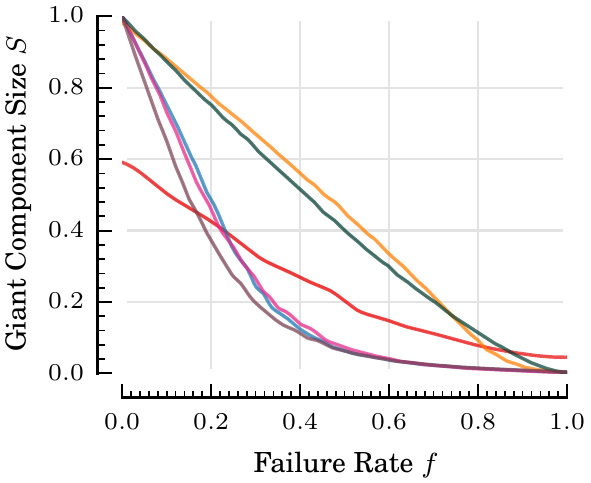}%
}%
\hspace{0.8cm}
\subfloat[Temporal robustness]{\label{fig:4sys_rand:rel_temp_eff}%
\includegraphics[scale=\gfxscale]{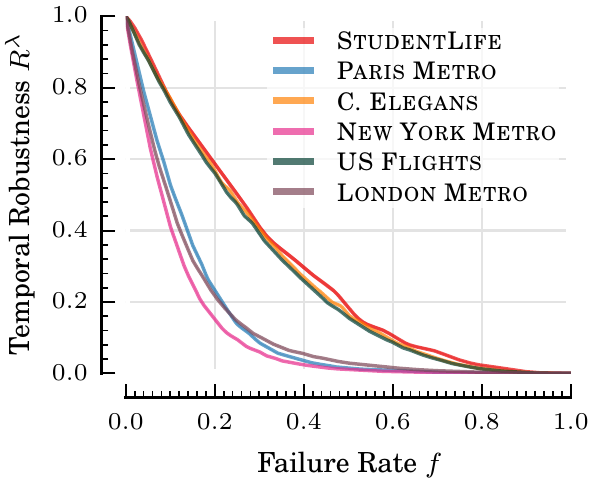}%
}%
\\
\subfloat[Spatial robustness]{\label{fig:4sys_rand:rel_spat_eff}%
\includegraphics[scale=\gfxscale]{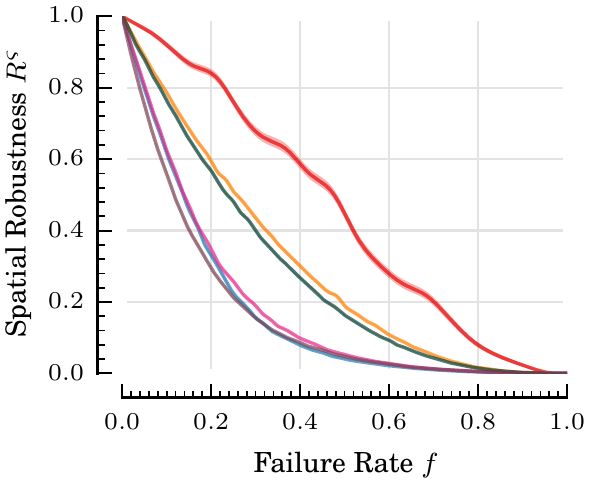}%
}%
\caption{\label{fig:4sys_rand}(Colour) Vulnerability to random node failure measured by giant strongly connected component size $S$, temporal robustness $\SrobTime$, spatial robustness $\SrobDist$. Expected values for quantities were obtained through simulation of 1,000 random realisations at each deactivation rate, and are represented by solid curves. Standard error (shaded region surrounding each curve) is negligible.}
\end{figure*}

Fig.~\ref{fig:4sys_rand} shows the response of each robustness measure with respect to uniform random failure. We used 1,000 random realisations at each deactivation rate to obtain expected values for each quantity.
The \DSunderground network is substantially less robust according to all three measures. In Fig.~\ref{fig:4sys_rand:giant_strong} we see that the network becomes highly fragmented after very few failures, and the giant component is effectively eliminated (filling less than 5\% of the network) at failure rates above 0.6. 

In an ideal configuration, deactivating one node should have minimum effect in disconnecting other nodes in the network. We see that \DSflights and \Dcelegans are much closer to this resilient behaviour. When comparing the edge density of the directed aggregate networks (0.9\% for \DSunderground, 0.7\% for \DSparis, 0.6\% for \DSnewyork, 4.4\% for \DSflights, 3.9\% for \Dcelegans, and 14.7\% for \Dstudentlife) this may in part be explained by more path redundancy in the high-density networks. Although the giant component size in the intact \Dstudentlife network is smaller, the profile of its relative decay is similar to that of \DSflights and \Dcelegans.
We also observe that \Dcelegans is typically more resilient than \DSflights, despite \Dcelegans having slightly lower density. %
\Dcelegans also exhibits increasing giant-component vulnerability in the $f=0.5$ to $0.8$ range, eventually becoming more fragmented than \DSflights. Its temporal and spatial robustness (Fig.~\ref{fig:4sys_rand:rel_temp_eff} and Fig.~\ref{fig:4sys_rand:rel_spat_eff}), on the other hand, do not exhibit the same behaviour.

Comparing Fig.~\ref{fig:4sys_rand:rel_temp_eff} and Fig.~\ref{fig:4sys_rand:rel_spat_eff}, we note that the spatial robustness of \DSflights, \Dcelegans, \DSunderground, and \DSparis follow a degradation pattern similar to their temporal robustness. In these systems, spatio-temporal paths are constrained by finite node-to-node propagation speeds. The propagation speeds in these networks are also heterogeneous, with the amount of diversity depending on the particular system; for example, longer track segments in \DSunderground and longer flight paths in \DSflights tend to have higher average speeds (flight duration also depends on atmospheric conditions), and synaptic transmission speeds in C. Elegans depend on the signalling mechanism. On the other hand, the homogeneous infinite propagation speeds in \Dstudentlife permit transmission to occur instantaneously, independent of the physical separation between individuals, and thus we see very different behaviour in this network with respect to spatial and temporal robustness (Fig.~\ref{fig:4sys_rand:rel_temp_eff} and Fig.~\ref{fig:4sys_rand:rel_spat_eff}).

\subsection{Static spatial paths vs. spatio-temporal paths}

\begin{figure*}
\centering
\subfloat[\Dunderground]{\label{fig:pathcomparison:und}%
\includegraphics[width=0.34\textwidth]{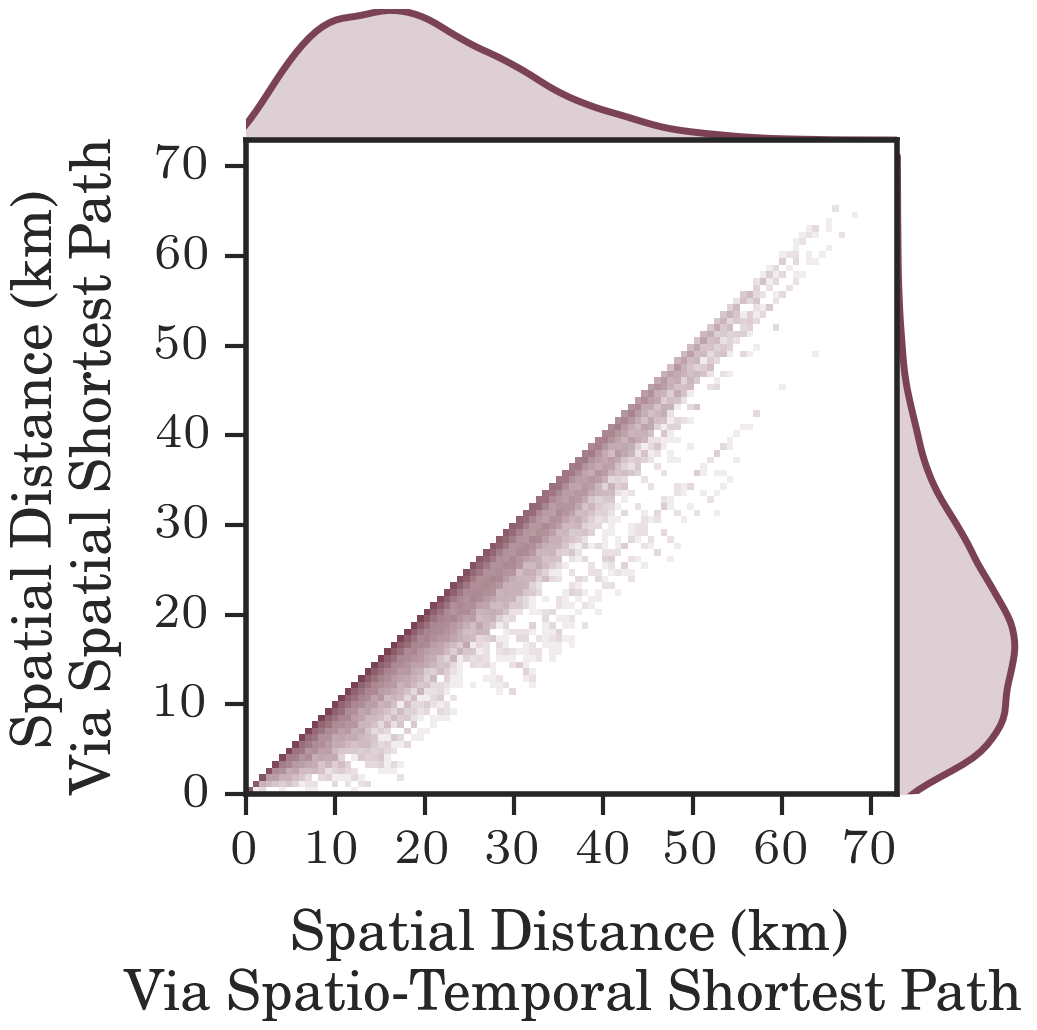}%
}%
\hspace{1.8cm} 
\subfloat[\Dparis]{\label{fig:pathcomparison:par}%
\includegraphics[width=0.34\textwidth]{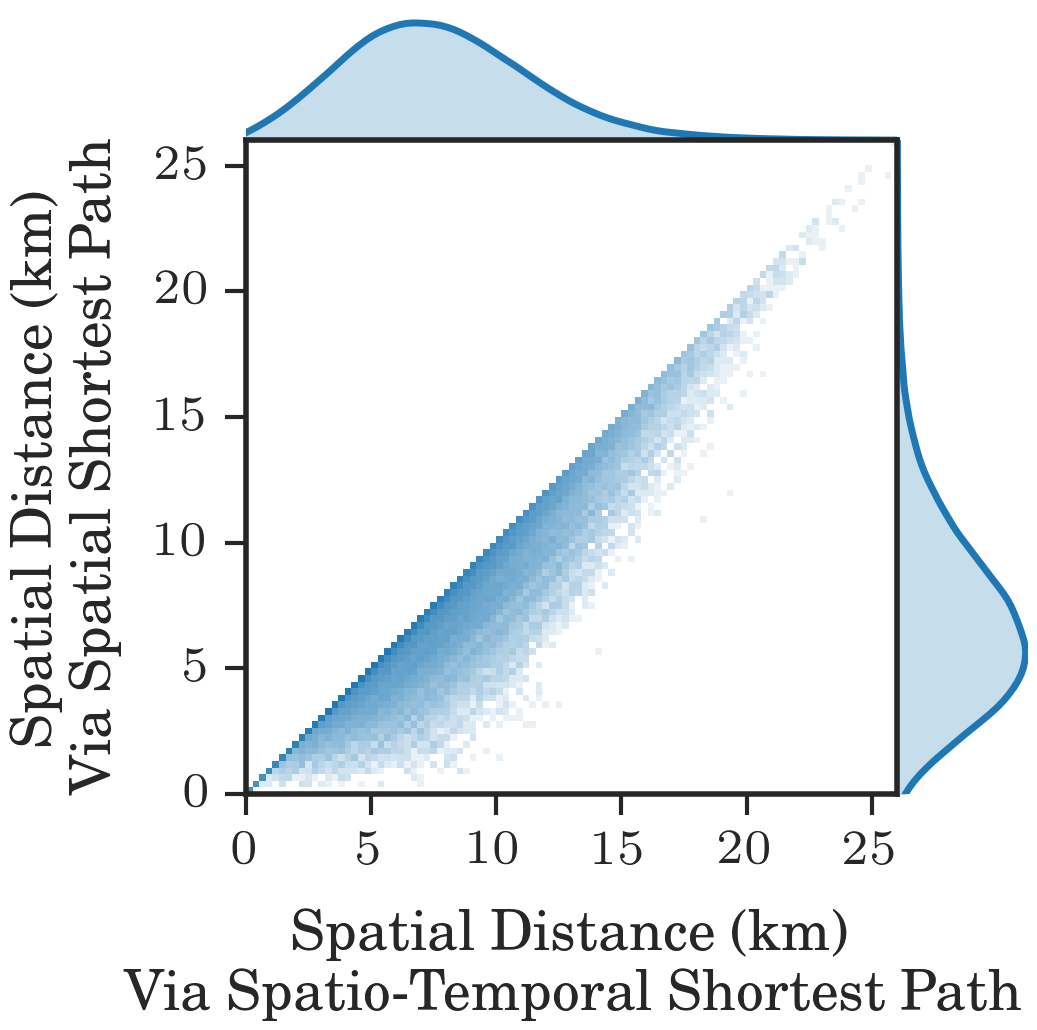}%
}%
\\
\subfloat[\Dnewyork]{\label{fig:pathcomparison:new}%
\includegraphics[width=0.34\textwidth]{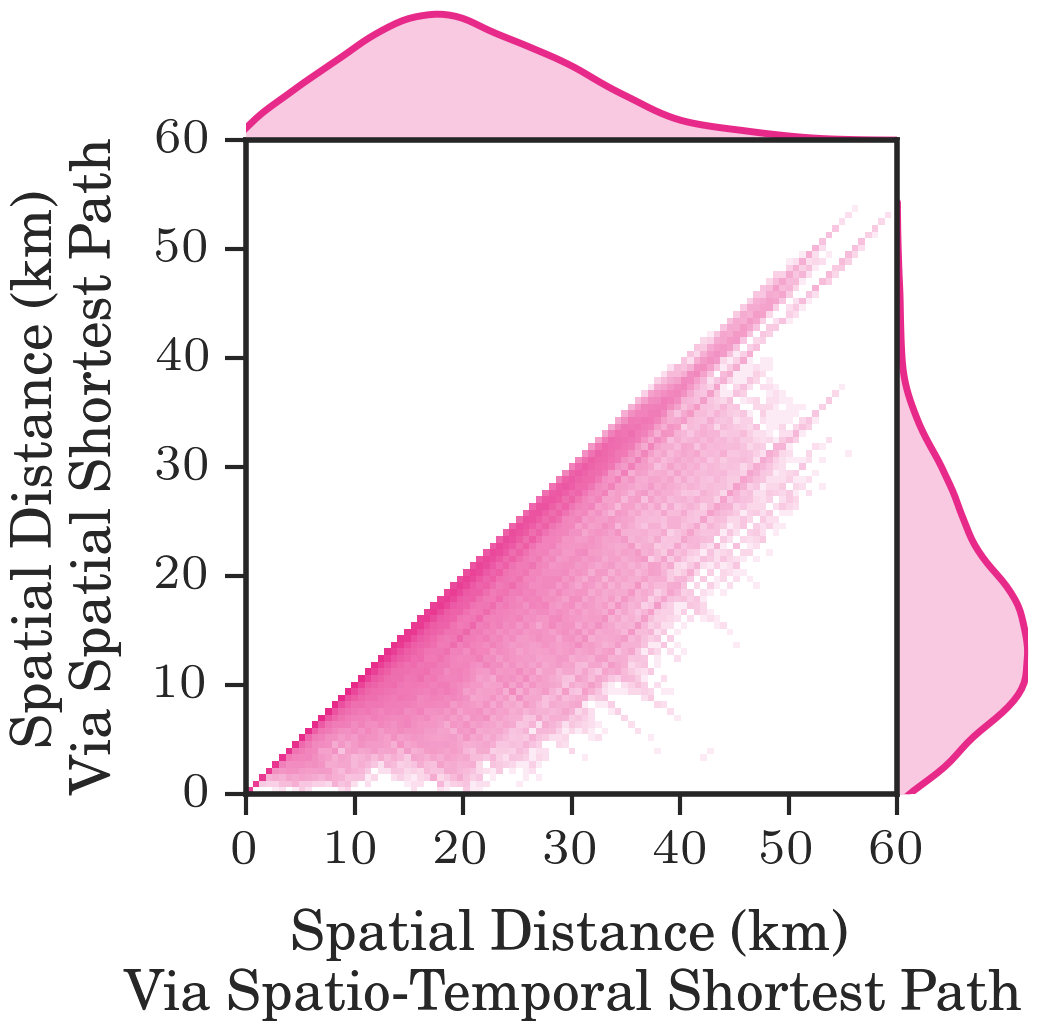}%
}%
\hspace{1.8cm} 
\subfloat[\Dflights]{\label{fig:pathcomparison:flights}%
\includegraphics[width=0.34\textwidth]{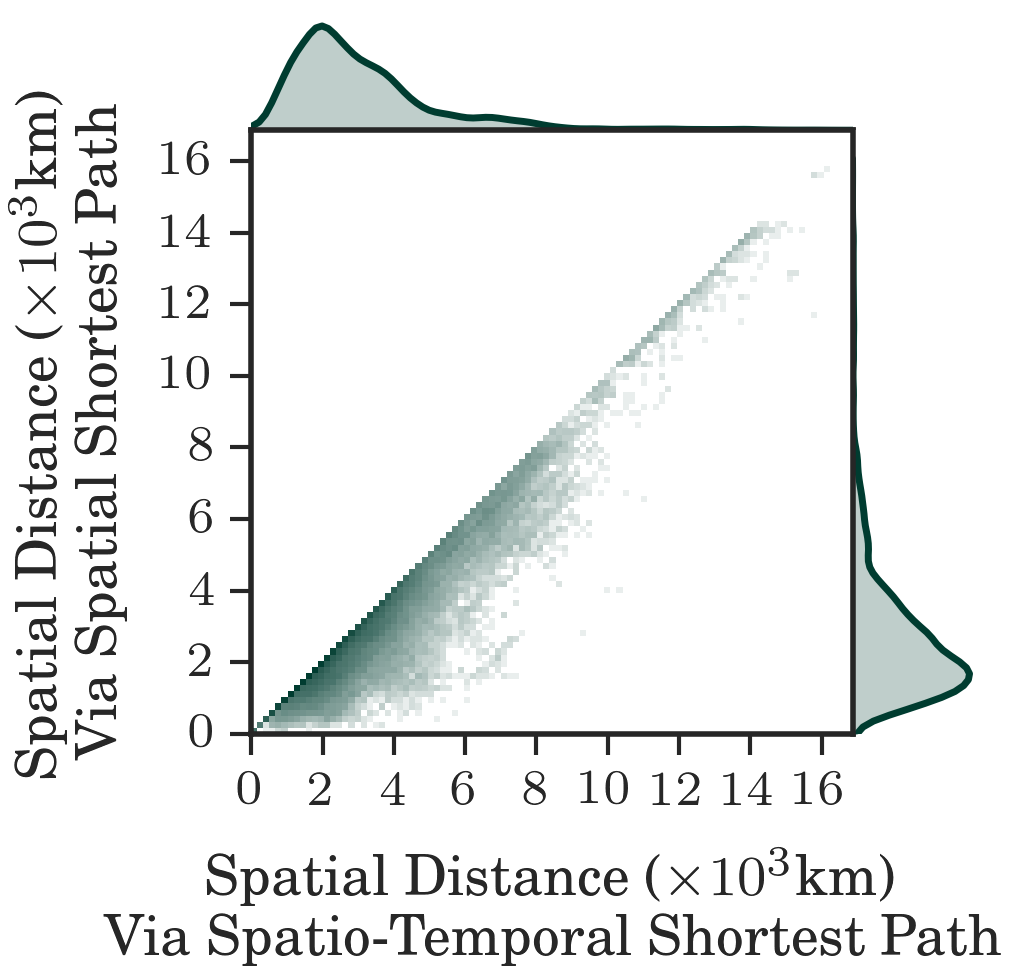}%
}%
\caption{\label{fig:pathcomparison}(Colour) Three urban transport networks and the US flights network. Comparison of spatio-temporal paths to shortest spatial paths in the aggregate spatial network, measured by spatial distance. Shortest spatial paths follow minimum spatial length. Each heatmap compares the spatial distances between pairs of stations according to definitions of spatial path and spatio-temporal path. Pearson correlation between these distances is 0.983, 0.952, 0.871, and 0.950 for \DSunderground, \DSparis, \DSnewyork, and \Dflights, respectively.}
\end{figure*}

Of the three urban transit systems, we note that \DSnewyork exhibits the greatest difference in response to random failure between spatial and temporal robustness (Fig.~\ref{fig:4sys_rand:rel_temp_eff} versus Fig.~\ref{fig:4sys_rand:rel_spat_eff}; area under curve comparison in Table~\ref{tab:attack_aucs}). 
In Fig~\ref{fig:pathcomparison}, we demonstrate the differing routes taken according to spatio-temporal paths versus shortest paths in the equivalent aggregate spatial network. We find that spatio-temporal distances tend to take longer routes through the network than purely spatial paths in \DSnewyork. Indeed, correlation between these pairwise distances is least in the \DSnewyork network; that is, minimum-latency routes tend to traverse longer distances in the \DSnewyork transit system. We note that the comparison in this figure is in the intact networks (i.e., without node deactivations), to highlight the differences in the types of paths in these systems.


\subsection{Tolerance to systematic attack}

To study attack tolerance we apply the deactivation strategies defined in Sec.~\ref{sec:attacks} to each network, allowing us to measure network performance after a fraction of nodes $f$ are deactivated. 

\begin{figure*}
\centering
\subfloat[\Dunderground]{\label{fig:gc:und}%
\includegraphics[scale=\gfxscale]{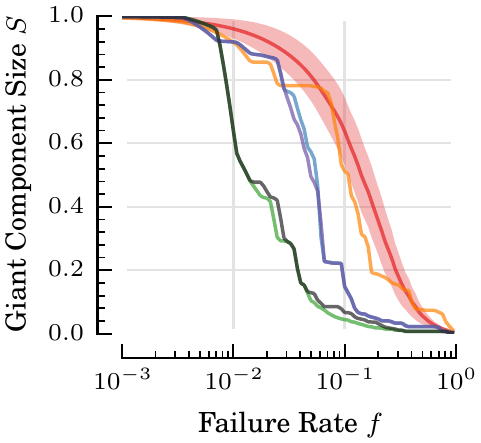}%
}%
\hspace{0.3cm}%
\subfloat[\Dparis]{\label{fig:gc:par}%
\includegraphics[scale=\gfxscale]{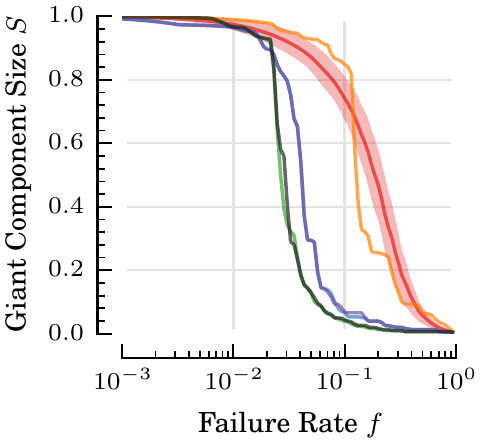}%
}%
\hspace{0.3cm}%
\subfloat[\Dnewyork]{\label{fig:gc:new}%
\includegraphics[scale=\gfxscale]{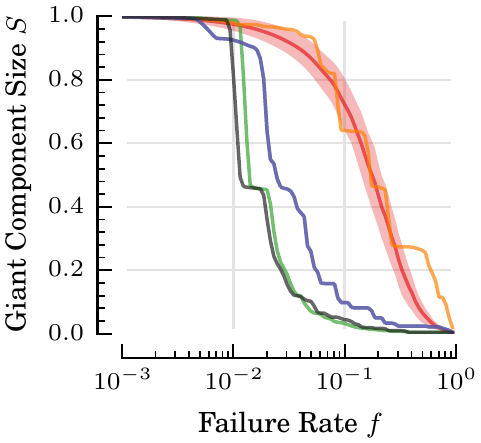}%
}%
\\\vspace{0.1cm}%
\subfloat[\Dflights]{\label{fig:gc:fli}%
\includegraphics[scale=\gfxscale]{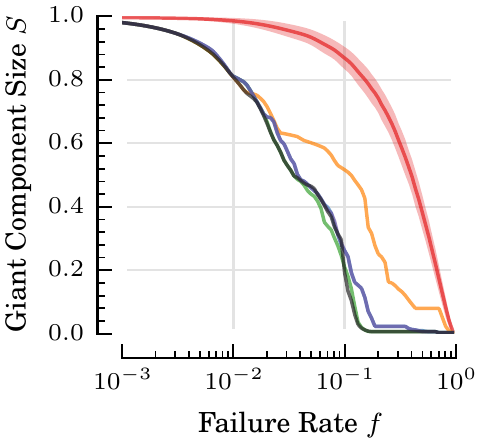}%
}%
\hspace{0.3cm}%
\subfloat[\Dcelegans]{\label{fig:gc:cel}%
\includegraphics[scale=\gfxscale]{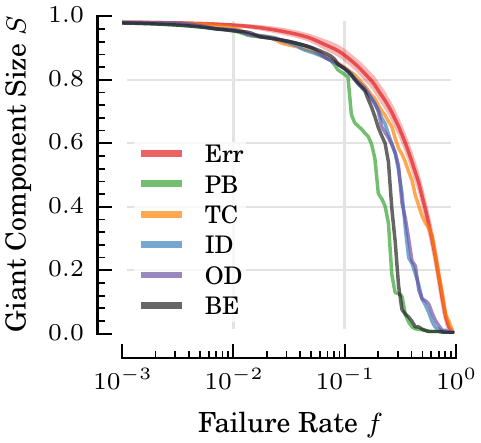}%
}%
\hspace{0.3cm}%
\subfloat[\Dstudentlife]{\label{fig:gc:std}%
\includegraphics[scale=\gfxscale]{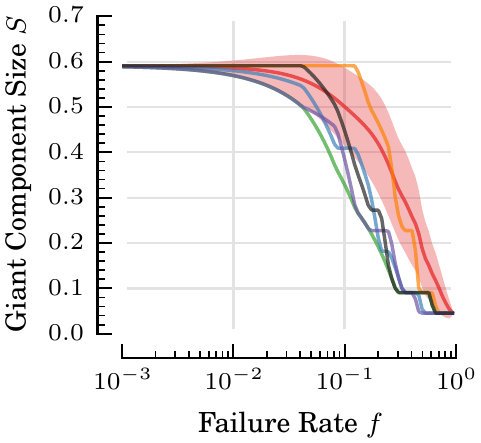}%
}%
\caption{\label{fig:gc} (Colour) Robustness of the giant strongly connected component when each system is subject to five systematic attack strategies: temporal closeness (TC), path betweenness (PB), betweenness efficiency (BE), in-degree (ID), and out-degree (OD). Vulnerability to random failure (Err) included for comparison. Shaded regions represent standard deviation (see Fig.~\ref{fig:4sys_rand} for standard error).}
\end{figure*}

\begin{figure*}
\centering
\subfloat[\Dunderground]{\label{fig:tr:und}%
\includegraphics[scale=\gfxscale]{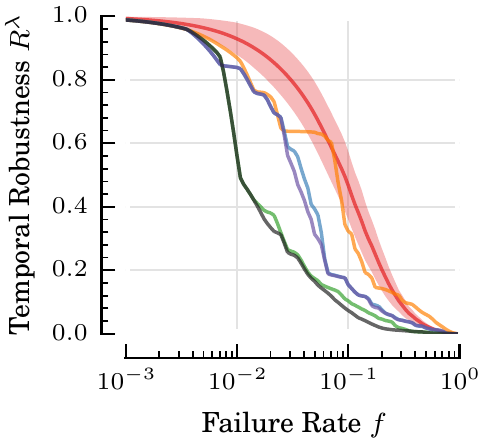}%
}%
\hspace{0.3cm}%
\subfloat[\Dparis]{\label{fig:tr:par}%
\includegraphics[scale=\gfxscale]{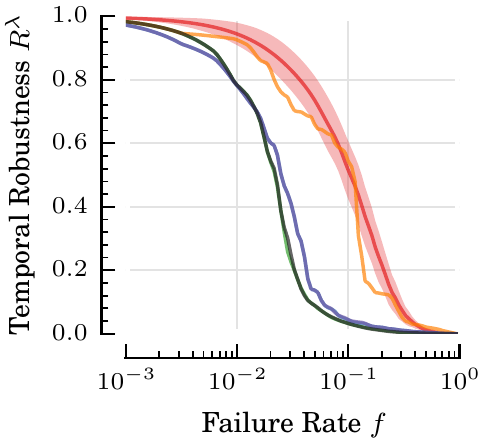}%
}%
\hspace{0.3cm}%
\subfloat[\Dnewyork]{\label{fig:tr:new}%
\includegraphics[scale=\gfxscale]{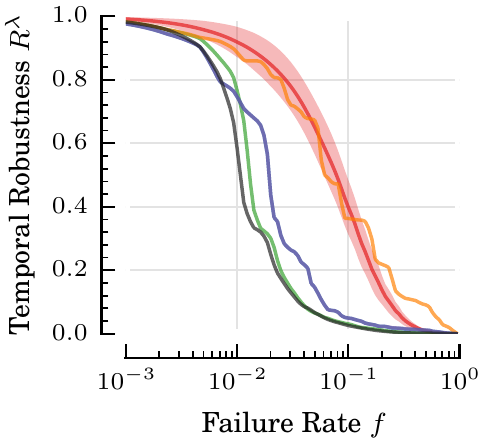}%
}%
\\\vspace{0.1cm}%
\subfloat[\Dflights]{\label{fig:tr:fli}%
\includegraphics[scale=\gfxscale]{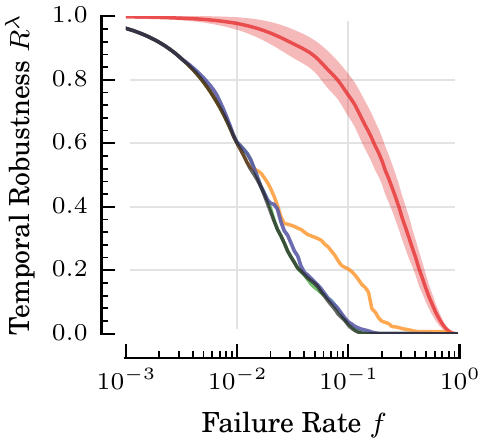}%
}%
\hspace{0.3cm}%
\subfloat[\Dcelegans]{\label{fig:tr:cel}%
\includegraphics[scale=\gfxscale]{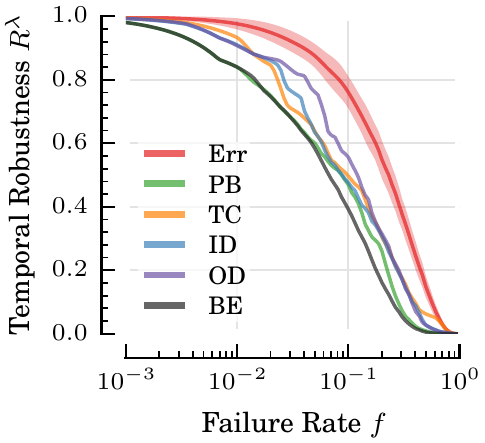}%
}%
\hspace{0.3cm}%
\subfloat[\Dstudentlife]{\label{fig:tr:std}%
\includegraphics[scale=\gfxscale]{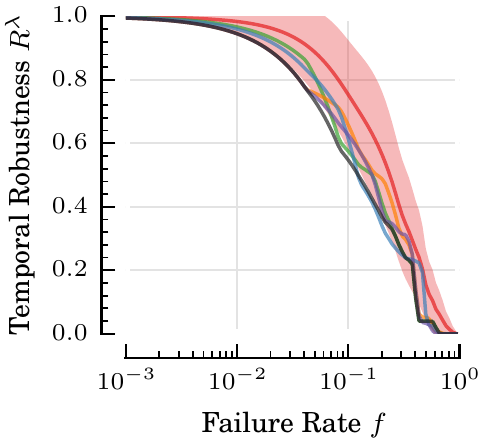}%
}%
\caption{\label{fig:tr} (Colour) Robustness of temporal efficiency under systematic attack. Labels and colours same as Fig.~\ref{fig:gc}. Shaded regions represent standard deviation (see Fig.~\ref{fig:4sys_rand} for standard error).}
\end{figure*}

Although all attack schemes are generally more effective than random error, there is substantial variation in their performance. In Fig.~\ref{fig:gc} we plot the effectiveness of each strategy at dismantling the giant component.
In \DSflights, the damage caused by the four non-closeness attacks is very similar. We see some slight variation in giant-component vulnerability with respect to each attack (Fig.~\ref{fig:gc:fli}), and very similar behaviour in temporal vulnerability (Fig.~\ref{fig:tr:fli}). This contrasts with \DSunderground and \Dcelegans, where the betweenness-based attacks follow substantially different profiles to the degree-based attacks.

The two betweenness attacks are very effective at dismantling the giant component in \DSunderground up to $f=0.05$ (Fig.~\ref{fig:gc:und}). At this point (i.e., with 5\% of the nodes deactivated), the nodes vital to connecting London's central core to peripheral Underground lines have been deactivated. A substantial proportion of nodes (roughly 190 of 270 stations) belong to isolated clusters in this now-disconnected outer region. There is more tolerance to failure in the remaining core, as we observe from the gradual decline in $S$ from $f=0.05$, and the giant component is finally eliminated at $f=0.45$. Interestingly, the threshold at which systematic attack eliminates the giant component in \DSflights (see Fig.~\ref{fig:gc:fli}) is much lower; specifically, this occurs at $f=0.16$ for \DSflights and at $f=0.45$ for \DSunderground.

Comparing Fig.~\ref{fig:gc} and Fig.~\ref{fig:tr}, we observe how the two betweenness centralities attack reachability and temporal efficiency differently. Path betweenness (PB) preferentially attacks nodes through which many geodesic paths flow, whereas betweenness efficiency (BE) prioritises nodes supporting temporally efficient paths. Thus, the giant component is harmed more rapidly in the PB attack (Fig.~\ref{fig:gc}), and temporal robustness declines more quickly under the BE attack (Fig.~\ref{fig:tr}). This difference is especially pronounced in \Dcelegans (Fig.~\ref{fig:gc:cel} and Fig.~\ref{fig:tr:cel}). 

To judge the overall effectiveness of a particular attack for a given performance measure, we also consider the area under each robustness curve. Values for random failure and the two betweenness-based attacks are provided in Table~\ref{tab:attack_aucs}. 
We see that these two attacks indeed preferentially degrade different features of a network. Specifically, Table~\ref{tab:attack_aucs} highlights that for \DSunderground, \Dcelegans, and \Dstudentlife, the lowest overall $S$-robustness is achieved with the PB attack strategy and the lowest overall $\SrobTime$-robustness is achieved with the BE attack strategy. We also find that in the other three networks (\DSflights, \DSnewyork, and \DSparis), the effect is much less pronounced, where both PB and BE have roughly the same effect with respect to the two vulnerability measures ($S$ and $\SrobTime$).
\begin{table}
\caption{\label{tab:attack_aucs}%
Overall effectiveness of path betweenness attack (PB) and betweenness efficiency attack (BE) measured by the area under each robustness curve. Areas are provided for two vulnerability measures: giant strong temporal component size ($S$) and temporal robustness ($\SrobTime$). A lower value indicates an attack that is more aggressive. Random failure (Err) is also included for comparison.
}
\centering
\newcommand{\xcolwid}{0.855cm}
\begin{tabular}{m{3.0cm}|C{\xcolwid}C{\xcolwid}C{\xcolwid}|C{\xcolwid}C{\xcolwid}C{\xcolwid}}
\cline{1-7}
            & \multicolumn{6}{c}{Area Under Robustness Curve} \\
\cline{2-7}
            & \multicolumn{3}{c|}{$S$} & \multicolumn{3}{c}{$\SrobTime$} \\
              &  Err  &  PB    &  BE       &  Err     &  PB      &  BE      \\
\hline
\Dunderground  & .195  & \textbf{.035}   & .039      & .134     & .042     & \textbf{.034}     \\
\Dparis        & .224  & \textbf{.043}   & \textbf{.043}      & .136     & \textbf{.030}     & \textbf{.030}     \\
\Dnewyork      & .221  & \textbf{.031}   & \textbf{.031}      & .110     & .024     & \textbf{.022}     \\
\Dflights      & .431  & \textbf{.060}   & \textbf{.060}      & .274     & \textbf{.025}     & \textbf{.025}     \\
\Dcelegans     & .448  & \textbf{.201}   & .234      & .278     & .124     & \textbf{.106}     \\
\Dstudentlife  & .244  & \textbf{.132}   & .152      & .287     & .202     & \textbf{.190}     \\
\hline
\end{tabular}
%
\end{table}

In general, we observe that both betweenness-based attacks harm each network more effectively than the degree-based strategies. This is intuitive, as betweenness-based attacks are able to leverage global network information to decide their targets. 
What is more surprising is that temporal closeness, which is also a global-knowledge attack, is ineffective at identifying vulnerabilities in the network, and is often unable to out-perform the na\"ive degree-based attacks. The same is true for the strategy's effectiveness in attacking the network's temporal efficiency (Fig.~\ref{fig:tr}). 
Temporal (in-)closeness specifically identifies `sink' nodes that are effective endpoints for information flow. These results show that such sinks do not tend to be important centres on which other nodes rely, and therefore their deactivation has relatively little impact on the performance of the network.

We see that \Dcelegans is substantially more robust to systematic attack. Using the most-effective strategy in each network as a comparison, we see the giant component in \Dcelegans is highly resilient (Fig.~\ref{fig:gc:cel}), managing to remain at over 80\% coverage up to $f=0.12$, whereas other networks begin rapid degradation at $f=0.01$ or lower.

Finally, we compare the responses of the four transport networks to the most-aggressive attack strategies. We see that temporal and giant component degradation for \DSunderground (Fig.~\ref{fig:tr:und} and Fig.~\ref{fig:gc:und}, respectively) both follow very similar patterns, indicating that mutual reachability declines at a similar rate to the relative decline in temporal efficiency. This contrasts with \DSflights, whose giant component (Fig.~\ref{fig:gc:fli}) is more resilient than its temporal robustness (Fig.~\ref{fig:tr:fli}). This differing behaviour shows that there is a period during systematic attack on \DSflights where the network suffers increasing temporal delay while its giant component remains almost intact.

\subsection{\label{sec:centcorr}Attack strategy correlations}
\begin{figure*}
\centering
\subfloat[\Dunderground]{\label{fig:centcorr:und}%
\includegraphics[width=0.34\textwidth]{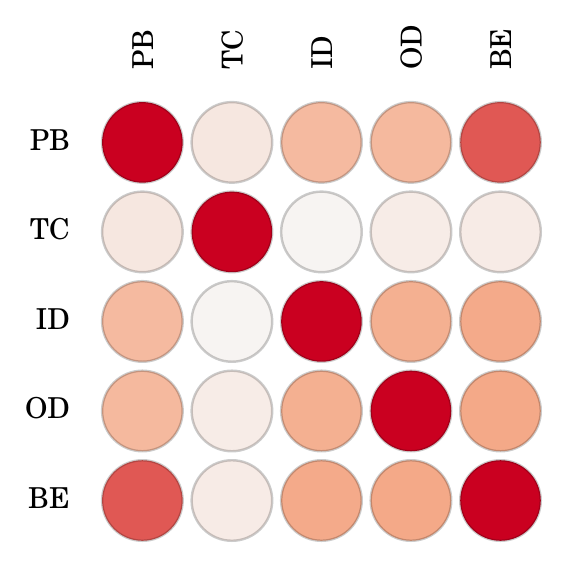}%
}%
\hspace{1.5cm} 
\subfloat[\Dparis]{\label{fig:centcorr:par}%
\includegraphics[width=0.34\textwidth]{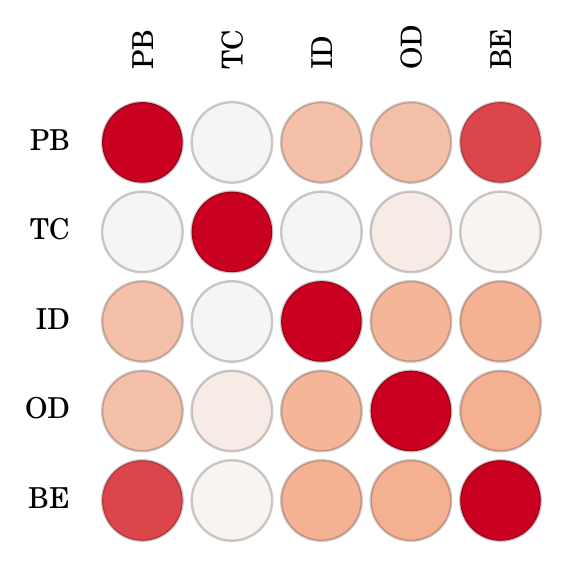}%
}%
\\
\subfloat[\Dnewyork]{\label{fig:centcorr:new}%
\includegraphics[width=0.34\textwidth]{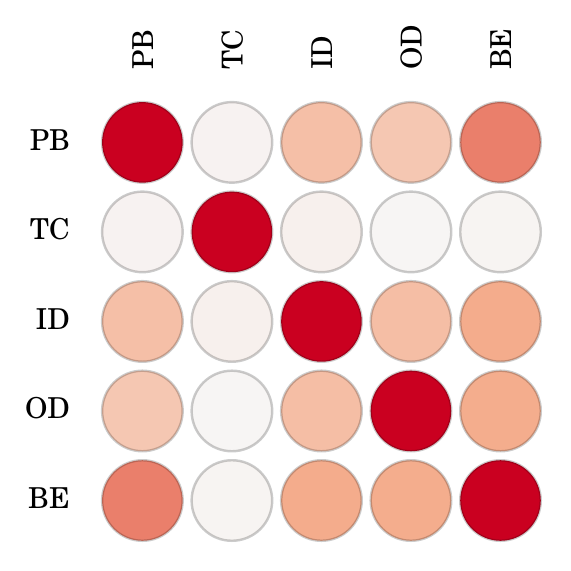}%
}%
\hspace{1.5cm} 
\subfloat[\Dflights]{\label{fig:centcorr:flights}%
\includegraphics[width=0.34\textwidth]{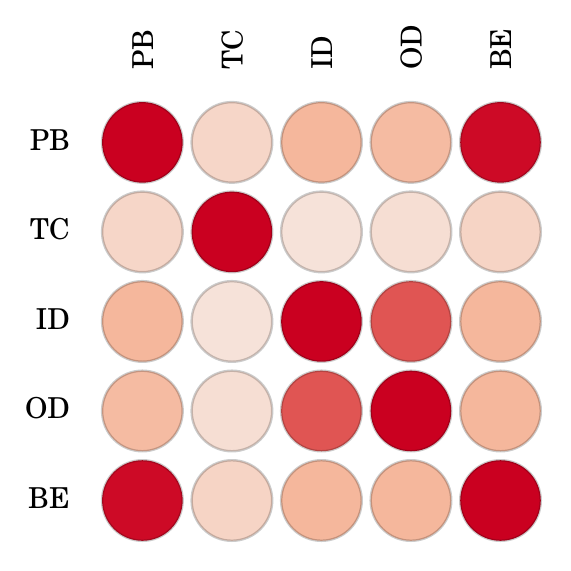}%
}%
\\
\vspace{0.4cm}
\includegraphics[width=0.28\textwidth]{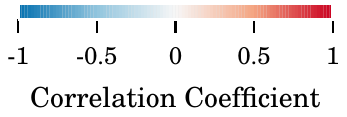}%
\caption{\label{fig:centcorr}(Colour) Three urban transport networks and the US flights network. Comparison of the node deactivation order according to each strategy in each network. Correlation is measured by the Kendall rank correlation coefficient (Kendall's tau).
Attack strategies as follows: path betweenness (PB), temporal closeness (TC), in degree (ID), out degree (OD), and betweenness efficiency (BE).}
\end{figure*}

To explore the attack strategies further, we compare the order with which each strategy preferentially deactivates nodes (Fig.~\ref{fig:centcorr}). Unsurprisingly, examples where we see the biggest difference in attack performance are also where we see lowest correlation in deactivation order.
For example, in \Dparis and \Dflights we see high correlation among most pairs of strategies, and also observe that these strategies have similar robustness profiles, especially in the case of \Dflights.
Comparing centrality correlation between the two betweenness attacks (BE and PB) also offers insight. These are overall highly correlated in all networks. Specifically, we observe correlation coefficients %
0.732 for \Dunderground, %
0.788 for \Dparis, %
0.613 for \Dnewyork, %
0.961 for \Dflights, and %
0.613 for \Dcelegans (not plotted). However, where we see less agreement between these two strategies, we also see the greatest differences in their effectiveness at attacking the giant component (Fig.~\ref{fig:gc}) versus temporal robustness (Fig.~\ref{fig:tr}). This effect is most pronounced in \Dcelegans and \Dnewyork, which also have the lowest PB-BE correlation of the networks we studied.


\subsection{Comparison with temporal centrality}

\begin{figure*}
\centering
\subfloat[\Dunderground]{\label{a1}%
\includegraphics[width=0.34\textwidth]{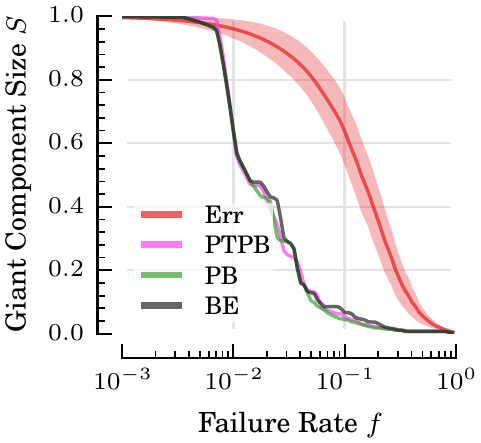}%
\hspace{1.5cm} 
\includegraphics[width=0.34\textwidth]{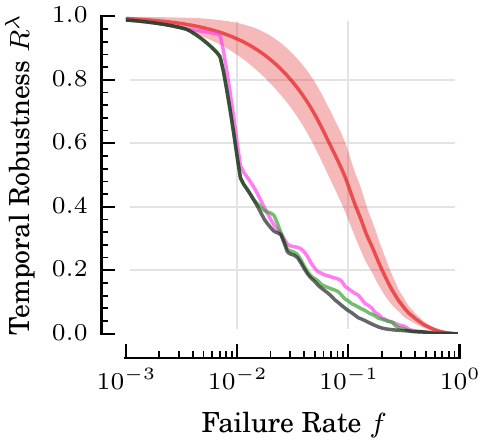}%
}%
\\
\subfloat[\Dnewyork]{\label{a3}%
\includegraphics[width=0.34\textwidth]{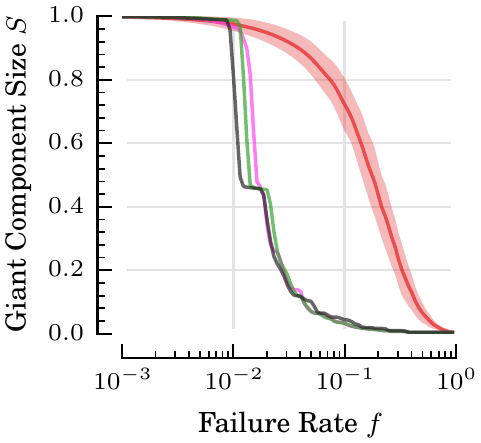}%
\hspace{1.5cm} 
\includegraphics[width=0.34\textwidth]{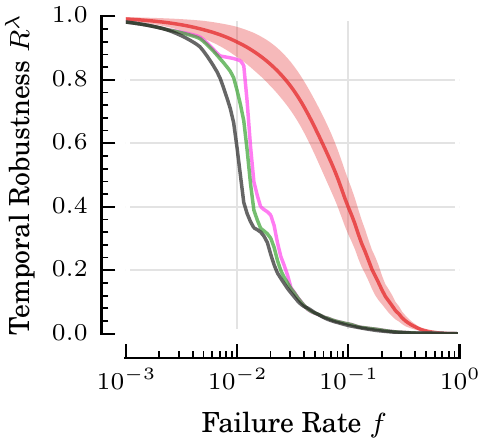}%
}%
\caption{\label{fig:pure_temp_comparison}(Colour) Comparison of spatio-temporal attacks (PB and BE) and the pure-temporal path betweenness attack (PTPB).}
\end{figure*}

Finally, we compare the two attack strategies based on spatio-temporal paths (PB and BE) to a purely temporal counterpart. Here we explore whether a centrality measure based on temporally shortest paths has the same effect as PB and BE. Specifically, we apply the temporal betweenness centrality introduced in~\cite{nicosia:2013:graph}, which for convenience we refer to as PTPB (pure temporal path betweenness). In the calculation of temporal betweeneess centrality, a node's betweenness is counted as the number of shortest temporal paths through it, as opposed to the number of shortest spatio-temporal paths. In other words, these temporal paths are not constrained by distance and propagation speed. (Equivalently, we can regard this as the case where all propagation speeds are infinite.) 
We show the performance of these attacks for two metro systems, \Dunderground and \Dnewyork, in Fig~\ref{fig:pure_temp_comparison}. We see that the space-agnostic attack is less effective than PB and BE in both systems.




\section{\label{sec:conclusions}Conclusions}

The ability to explicitly represent non-instantaneous propagation that reflects the space and time in which a network is embedded allows us to better understand the properties of real-world spatio-temporal systems. In this paper, we develop a framework that combines these two dimensions, constructing networks whose connectivity is constrained by the dynamics of the real-world systems they represent. With this framework we explore the resilience of these systems in terms of their topological, spatial, and temporal behaviour. 
The proposed approach is based on a constrained-propagation model over a time-varying representation of interaction speeds between nodes, and is a re-formulation of previous discrete-time instantaneous-spreading models of temporal networks.
Through numerical experiments on empirical systems we study the spatial, temporal, and topological performance of spatio-temporal networks when subject to random error and systematic attack.
Our approach is particularly motivated by the joint spatial and temporal nature of urban transport systems, while we additionally also test the generality of the approach on other classes of network.
We introduce global- and local-knowledge attack strategies, finding that attacks based on the betweenness of nodes perform better than other strategies. Most saliently, we find that the spatial, temporal, and topological responses of a system to node failure can behave differently to the same attack strategy. In particular, we find that systematic attack based on betweenness centrality is typically most effective at harming the temporal efficiency of a network, leading to increased delays, whereas path betweenness is most effective at eliminating mutual reachability within a network.


\vskip6pt

%
%
%

\section*{Acknowledgements}
The authors would like to thank Fani Tsapeli and Luca Canzian for their help in preparing the StudentLife mobility data, 
Petra Vertes for insights in reconstructing the C. Elegans connectome, Manuel Zimmer for advice on synaptic delay, 
and Luca Rossi for fruitful discussions.
Data processing and numerical experiments were performed in part using the computational facilities of the University of Birmingham's BEAR division.

\section*{Data accessibility}
Datasets~\cite{williams:2016:data} used in this research are available via the Dryad Repository (\url{http://dx.doi.org/10.5061/dryad.3p27r} -- doi:10.5061/dryad.3p27r). Visualisations~\cite{williams:2016:visualisations} of these networks are available via the Figshare Repository (\url{https://dx.doi.org/10.6084/m9.figshare.1452976} -- doi:10.6084/m9.figshare.1452976). These datasets and visualisations are further described in this paper's Electronic Supplementary Material.

\section*{Authors' contributions}
M.J.W. and M.M. designed and developed the research, performed the analyses, reviewed the results, and wrote the manuscript.

\section*{Competing interests}
The authors declare that they have no competing interests.

\section*{Funding statement}
This work was supported by \emph{LASAGNE} (Multi-layer Spatiotemporal Generalized Networks), a European Commission STREP project (grant ref. 318132), and \emph{The Uncertainty of Identity}, an EPSRC project (grant ref. EP/J005266/1).

\appendix

\section*{Appendices}

%
%
%

\section{\label{sec:reciprocity}Average reciprocity in a time-varying network}

The reciprocity of a (single) directed graph characterises the overall extent to which each edge in the network is in some way balanced by an edge in the opposite direction. This measure is commonly defined for binary graphs~\cite{garlaschelli:2004:patterns}, and has more recently been adapted to study weighted graphs~\cite{squartini:2013:reciprocity}. Here we describe how we extend this measure to characterise the average reciprocity in a time-varying weighted directed network represented by the sequence of weight matrices 
\begin{equation}
\Sadjmat^{[t_1]}, \, \Sadjmat^{[t_2]}, \, \ldots, \, \Sadjmat^{[t_T]} \,.
\end{equation}

First, let us consider a single weight matrix $\Sadjmat^{[t]}$ and obtain the global weighted reciprocity $\rho^{[t]}$ of this (static) graph as defined by Squartini et al~\cite{squartini:2013:reciprocity}. That is, given the total weight of the graph
\begin{equation}
W^{[t]} = \sum_v \sum_{w\not=v} \Sadjmat^{[t]}_{vw}
\end{equation}
and the total reciprocated weight of the graph
\begin{equation}
W^{[t]}_\leftrightarrow = \sum_v \sum_{w\not=v}
  \mathrm{min}( \Sadjmat^{[t]}_{vw}, \, \Sadjmat^{[t]}_{wv} )  \,,
\end{equation}
we obtain the global weighted reciprocity as
\begin{equation}
\rho^{[t]} = \frac{W^{[t]}_\leftrightarrow}{W^{[t]}} \,.
\end{equation}
In the case that the graph is empty, we set $\rho^{[t]} = 1$. Finally, as a diagnostic of the reciprocity in the whole time-varying network, we define the \emph{weight reciprocity} as the average weighted reciprocity
\begin{equation}
	\SrecipW = \frac{1}{T} \sum_{i=1}^{T} \rho^{[t_i]} \, .
\end{equation}
Weight reciprocity is normalised between 0 and 1. $\SrecipW=1$ indicates that each edge in the network is reciprocated by an identically weighted edge in the opposite direction, and thus the network is effectively undirected. In the opposite case, where each edge in the network has no reciprocal counterpart, we have $\SrecipW=0$. We note that for the trivial case of an empty (edgeless) time-varying network we have $\SrecipW=1$.

It is sometimes useful to ignore weights in the time-varying network and study reciprocity from a purely topological viewpoint. We refer to this as the (average) \emph{topological reciprocity}, denoted by $\SrecipT$. This quantity is the average reciprocity over the network's sequence of binary adjacency matrices.

%
%
%

\section{\label{sec:gyration}Radius of gyration}

To characterise the typical physical distance travelled by a node in a spatio-temporal system we use the average radius of gyration taken over all nodes in the system. We follow the same approach as previous work that has adapted the classical mechanical definition of radius of gyration to discrete-time trajectories of moving objects~\cite{gonzalez:2008:understanding}.

Recall that in our representation of a spatio-temporal system, nodes are embedded in a $k$-dimensional metric space with physical distance function $g$. Given a node $v$ and its trajectory represented by the time-varying sequence of positions $l_v^{[1]}, \, l_v^{[2]}, \, \ldots, \, l_v^{[T]}$ the radius of gyration of $v$ is expressed as
\begin{equation}
\Srofg^v = 
\left(
  \frac{1}{T}
  \sum_{i=1}^{T} \, g(l_v^{[i]}, \, \Scofm^v)^2
\right)^{1/2}
\end{equation}
where $\Scofm^v$ is the centroid of the trajectory of $v$.
Our summary of the mobility of the set of all nodes $V$ in a system is then given by the average radius of gyration
\begin{equation}
\SrofgAvg = \frac{1}{|V|}
\, \sum_{v \in V} \, \Srofg^v \, .
\end{equation}





\bibliographystyle{plain}

\bibliography{refs}

\begin{thebibliography}{10}

\bibitem{agarwal:2013:resilience}
Pankaj~K. Agarwal, Alon Efrat, Shashidhara~K. Ganjugunte, David Hay,
  Swaminathan Sankararaman, and Gil Zussman.
\newblock The {Resilience} of {WDM} {Networks} to {Probabilistic}
  {Geographical} {Failures}.
\newblock {\em IEEE/ACM Trans. Netw.}, 21(5):1525--1538, October 2013.

\bibitem{albert:2004:structural}
R.~Albert, I.~Albert, and G.~L. Nakarado.
\newblock Structural {Vulnerability} of the {North} {American} {Power} {Grid}.
\newblock {\em Phys. Rev. E}, 69(2):025103, February 2004.

\bibitem{albert:2000:error}
R.~Albert, H.~Jeong, and A.-L. Barab{\'a}si.
\newblock Error and {Attack} {Tolerance} of {Complex} {Networks}.
\newblock {\em Nature}, 406(6794):378--382, July 2000.

\bibitem{alsayed:2015:betweenness}
A.~Alsayed and D.~J. Higham.
\newblock Betweenness in {Time} {Dependent} {Networks}.
\newblock {\em Chaos, Solitons Fractals}, 72:35--48, March 2015.

\bibitem{angeloudis:2006:large}
Panagiotis Angeloudis and David Fisk.
\newblock Large {Subway} {Systems} as {Complex} {Networks}.
\newblock {\em Physica A}, 367:553--558, July 2006.

\bibitem{bajardi:2011:dynamical}
P.~Bajardi, A.~Barrat, F.~Natale, L.~Savini, and V.~Colizza.
\newblock Dynamical {Patterns} of {Cattle} {Trade} {Movements}.
\newblock {\em PLoS ONE}, 6(5):E19869, May 2011.

\bibitem{barthelemy:2011:spatial}
M.~Barth{\'e}lemy.
\newblock Spatial {Networks}.
\newblock {\em Phys. Rep.}, 499(1{\textendash}3):1--101, February 2011.

\bibitem{berche:2009:resilience}
B.~Berche, C.~von Ferber, T.~Holovatch, and Y.~Holovatch.
\newblock Resilience of {Public} {Transport} {Networks} {Against} {Attacks}.
\newblock {\em Eur. Phys. J. B}, 71(1):125--137, August 2009.

\bibitem{berezin:2015:localized}
Y.~Berezin, A.~Bashan, M.~M. Danziger, D.~Li, and S.~Havlin.
\newblock Localized {Attacks} on {Spatially} {Embedded} {Networks} with
  {Dependencies}.
\newblock {\em Sci. Rep.}, 5:08934, March 2015.

\bibitem{boccaletti:2014:structure}
S.~Boccaletti, G.~Bianconi, R.~Criado, C.~I. del Genio,
  J.~G{\'o}mez-Garde{\~n}es, M.~Romance, I.~Sendi{\~n}a-Nadal, Z.~Wang, and
  M.~Zanin.
\newblock The {Structure} and {Dynamics} of {Multilayer} {Networks}.
\newblock {\em Phys. Rep.}, 544(1):1--122, November 2014.

\bibitem{bota:2015:architecture}
M.~Bota, O.~Sporns, and L.~W. Swanson.
\newblock Architecture of the {Cerebral} {Cortical} {Association} {Connectome}
  {Underlying} {Cognition}.
\newblock {\em Proc. Natl. Acad. Sci. USA}, 112(16):E2093--E2101, April 2015.

\bibitem{brandes:2001:faster}
U.~Brandes.
\newblock A {Faster} {Algorithm} for {Betweenness} {Centrality}.
\newblock {\em The Journal of Mathematical Sociology}, 25(2):163--177, June
  2001.

\bibitem{buldyrev:2010:catastrophic}
S.~V. Buldyrev, R.~Parshani, G.~Paul, H.~E. Stanley, and S.~Havlin.
\newblock Catastrophic {Cascade} of {Failures} in {Interdependent} {Networks}.
\newblock {\em Nature}, 464(7291):1025--1028, April 2010.

\bibitem{callaway:2000:network}
D.~S. Callaway, M.~E.~J. Newman, S.~H. Strogatz, and D.~J. Watts.
\newblock Network {Robustness} and {Fragility}: {Percolation} on {Random}
  {Graphs}.
\newblock {\em Phys. Rev. Lett.}, 85(25):5468--5471, December 2000.

\bibitem{chen:2006:wiring}
B.~L. Chen, D.~H. Hall, and D.~B. Chklovskii.
\newblock Wiring {Optimization} {Can} {Relate} {Neuronal} {Structure} and
  {Function}.
\newblock {\em Proc. Natl. Acad. Sci. USA}, 103(12):4723--4728, March 2006.

\bibitem{cohen:2000:resilience}
R.~Cohen, K.~Erez, D.~ben Avraham, and S.~Havlin.
\newblock Resilience of the {Internet} to {Random} {Breakdowns}.
\newblock {\em Phys. Rev. Lett.}, 85(21):4626--4628, November 2000.

\bibitem{dallasta:2006:vulnerability}
L.~Dall{\textquoteright}Asta, A.~Barrat, M.~Barth{\'e}lemy, and A.~Vespignani.
\newblock Vulnerability of {Weighted} {Networks}.
\newblock {\em J. Stat. Mech. Theor. Exp.}, 2006(04):P04006, April 2006.

\bibitem{de_domenico:2014:navigability}
M.~de~Domenico, A.~Sol{\'e}-Ribalta, S.~G{\'o}mez, and A.~Arenas.
\newblock Navigability of {Interconnected} {Networks} {Under} {Random}
  {Failures}.
\newblock {\em Proc. Natl. Acad. Sci. USA}, 111(23):8351--8356, June 2014.

\bibitem{derrible:2010:complexity}
Sybil Derrible and Christopher Kennedy.
\newblock The {Complexity} and {Robustness} of {Metro} {Networks}.
\newblock {\em Physica A}, 389(17):3678--3691, September 2010.

\bibitem{dorbritz:2009:stability}
R.~Dorbritz and U.~Weidmann.
\newblock Stability of {Public} {Transportation} {Systems} in {Case} of
  {Random} {Failures} and {Intended} {Attacks} {\textendash} {A} {Case} {Study}
  from {Switzerland}.
\newblock In {\em Proc. 4th {IET} {International} {Conference} on {System}
  {Safety}, {Incorporating} the {SaRS} {Annual} {Conference}}, pages 1--6,
  October 2009.

\bibitem{eagle:2009:inferring}
N.~Eagle, A.~Pentland, and D.~Lazer.
\newblock Inferring {Friendship} {Network} {Structure} by using {Mobile}
  {Phone} {Data}.
\newblock {\em Proc. Natl. Acad. Sci. USA}, 106(36):15274--15278, 2009.

\bibitem{fenu:2015:block}
Caterina Fenu and Desmond~J. Higham.
\newblock Block {Matrix} {Formulations} for {Evolving} {Networks}.
\newblock {\em arXiv:1511.07305 [physics]}, November 2015.
\newblock arXiv: 1511.07305.

\bibitem{fornito:2015:connectomics}
A.~Fornito, A.~Zalesky, and M.~Breakspear.
\newblock The {Connectomics} of {Brain} {Disorders}.
\newblock {\em Nat. Rev. Neurosci.}, 16(3):159--172, March 2015.

\bibitem{gallotti:2015:multilayer}
R.~Gallotti and M.~Barth{\'e}lemy.
\newblock The {Multilayer} {Temporal} {Network} of {Public} {Transport} in
  {Great} {Britain}.
\newblock {\em Sci. Data}, 2:140056, January 2015.

\bibitem{gallotti:2014:anatomy}
Riccardo Gallotti and Marc Barthelemy.
\newblock Anatomy and {Efficiency} of {Urban} {Multimodal} {Mobility}.
\newblock {\em Sci. Rep.}, 4, November 2014.

\bibitem{garlaschelli:2004:patterns}
D.~Garlaschelli and M.~I. Loffredo.
\newblock Patterns of {Link} {Reciprocity} in {Directed} {Networks}.
\newblock {\em Phys. Rev. Lett.}, 93(26):268701, December 2004.

\bibitem{gonzalez:2008:understanding}
M.~C. Gonzalez, C.~A. Hidalgo, and A.-L. Barab{\'a}si.
\newblock Understanding individual human mobility patterns.
\newblock {\em Nature}, 453(7196):779--782, 2008.

\bibitem{grindrod:2010:evolving}
P.~Grindrod and D.~J. Higham.
\newblock Evolving {Graphs}: {Dynamical} {Models}, {Inverse} {Problems} and
  {Propagation}.
\newblock {\em Proc. R. Soc. A}, 466(2115):753--770, March 2010.

\bibitem{helbing:2013:globally}
D.~Helbing.
\newblock Globally {Networked} {Risks} and {How} to {Respond}.
\newblock {\em Nature}, 497(7447):51--59, May 2013.

\bibitem{holme:2005:network}
P.~Holme.
\newblock Network {Reachability} of {Real}-world {Contact} {Sequences}.
\newblock {\em Phys. Rev. E}, 71(4):046119, April 2005.

\bibitem{holme:2014:analyzing}
P.~Holme.
\newblock Analyzing {Temporal} {Networks} in {Social} {Media}.
\newblock {\em Proc. IEEE}, 102(12):1922--1933, December 2014.

\bibitem{holme:2002:attack}
P.~Holme, B.~J. Kim, C.~N. Yoon, and S.~K. Han.
\newblock Attack {Vulnerability} of {Complex} {Networks}.
\newblock {\em Phys. Rev. E}, 65(5):056109, May 2002.

\bibitem{holme:2012:temporal}
P.~Holme and J.~Saram{\"a}ki.
\newblock Temporal networks.
\newblock {\em Physics Reports}, 519(3):97--125, 2012.

\bibitem{holme:2013:temporal}
P.~Holme and J.~Saram{\"a}ki.
\newblock Temporal {Networks} as a {Modeling} {Framework}.
\newblock In {\em Temporal {Networks}}, pages 1--14. Springer, 2013.

\bibitem{kaiser:2006:nonoptimal}
M.~Kaiser and C.~C. Hilgetag.
\newblock Nonoptimal {Component} {Placement}, but {Short} {Processing} {Paths},
  due to {Long}-{Distance} {Projections} in {Neural} {Systems}.
\newblock {\em PLoS Comput. Biol.}, 2(7):E95, July 2006.

\bibitem{kaluza:2010:complex}
P.~Kaluza, A.~K{\"o}lzsch, M.~T. Gastner, and B.~Blasius.
\newblock The {Complex} {Network} of {Global} {Cargo} {Ship} {Movements}.
\newblock {\em J. R. Soc. Interface}, 7(48):1093--1103, July 2010.

\bibitem{karsai:2011:small}
M.~Karsai, M.~Kivel{\"a}, R.~K. Pan, K.~Kaski, J.~Kert{\'e}sz, A.-L.
  Barab{\'a}si, and J.~Saram{\"a}ki.
\newblock Small but {Slow} {World}: {How} {Network} {Topology} and {Burstiness}
  {Slow} {Down} {Spreading}.
\newblock {\em Phys. Rev. E}, 83(2):025102, February 2011.

\bibitem{kempe:2000:connectivity}
D.~Kempe, J.~Kleinberg, and A.~Kumar.
\newblock Connectivity and {Inference} {Problems} for {Temporal} {Networks}.
\newblock In {\em Proc. 32nd {Symposium} on {Theory} of {Computing}}, {STOC}
  '00, pages 504--513. ACM, 2000.

\bibitem{kivela:2014:multilayer}
M.~Kivel{\"a}, A.~Arenas, M.~Barth{\'e}lemy, J.~P. Gleeson, Y.~Moreno, and
  M.~A. Porter.
\newblock Multilayer {Networks}.
\newblock {\em J. Complex Networks}, 2(3):203--271, September 2014.

\bibitem{kivela:2012:multiscale}
Mikko Kivel{\"a}, Raj~Kumar Pan, Kimmo Kaski, J{\'a}nos Kert{\'e}sz, Jari
  Saram{\"a}ki, and M{\'a}rton Karsai.
\newblock Multiscale {Analysis} of {Spreading} in a {Large} {Communication}
  {Network}.
\newblock {\em J. Stat. Mech.}, 2012(03):P03005, 2012.

\bibitem{konschake:2013:robustness}
M.~Konschake, H.~H.~K. Lentz, F.~J. Conraths, P.~H{\"o}vel, and T.~Selhorst.
\newblock On the {Robustness} of {In}- and {Out}-components in a {Temporal}
  {Network}.
\newblock {\em PLoS ONE}, 8(2):E55223, February 2013.

\bibitem{latora:2001:efficient}
V.~Latora and M.~Marchiori.
\newblock Efficient {Behavior} of {Small}-{World} {Networks}.
\newblock {\em Phys. Rev. Lett.}, 87(19):198701, October 2001.

\bibitem{latora:2003:economic}
V.~Latora and M.~Marchiori.
\newblock Economic {Small}-world {Behavior} in {Weighted} {Networks}.
\newblock {\em Eur. Phys. J. B}, 32(2):249--263, March 2003.

\bibitem{lee:2008:statistical}
Keumsook Lee, Woo-Sung Jung, Jong~Soo Park, and M.~Y. Choi.
\newblock Statistical {Analysis} of the {Metropolitan} {Seoul} {Subway}
  {System}: {Network} structure and {Passenger} {Flows}.
\newblock {\em Physica A}, 387(24):6231--6234, October 2008.

\bibitem{lindsay:2011:optogenetic}
T.~H. Lindsay, T.~R. Thiele, and S.~R. Lockery.
\newblock Optogenetic {Analysis} of {Synaptic} {Transmission} in the {Central}
  {Nervous} {System} of the {Nematode} {Caenorhabditis} {Elegans}.
\newblock {\em Nat. Commun.}, 2:306, May 2011.

\bibitem{neumayer:2011:assessing}
S.~Neumayer, G.~Zussman, R.~Cohen, and E.~Modiano.
\newblock Assessing the {Vulnerability} of the {Fiber} {Infrastructure} to
  {Disasters}.
\newblock {\em IEEE/ACM Trans. Netw.}, 19(6):1610--1623, December 2011.

\bibitem{neumayer:2015:geographic}
Sebastian Neumayer, Alon Efrat, and Eytan Modiano.
\newblock Geographic {Max}-flow and {Min}-cut {Under} a {Circular} {Disk}
  {Failure} {Model}.
\newblock {\em Comput. Netw.}, 77(C):117--127, February 2015.

\bibitem{newman:2006:structure}
M.~E.~J. Newman, A.-L. Barab{\'a}si, and D.~J. Watts.
\newblock {\em The {Structure} and {Dynamics} of {Networks}}.
\newblock Princeton University Press, 2006.

\bibitem{newman:2010:networks:}
Mark Newman.
\newblock {\em Networks: {An} {Introduction}}.
\newblock Oxford University Press, March 2010.

\bibitem{nicosia:2013:graph}
V.~Nicosia, J.~Tang, C.~Mascolo, M.~Musolesi, G.~Russo, and V.~Latora.
\newblock Graph metrics for temporal networks.
\newblock In {\em Temporal {Networks}}, pages 15--40. Springer, 2013.

\bibitem{nicosia:2012:components}
V.~Nicosia, J.~Tang, M.~Musolesi, G.~Russo, C.~Mascolo, and V.~Latora.
\newblock Components in {Time}-varying {Graphs}.
\newblock {\em Chaos}, 22(2):023101, 2012.

\bibitem{nicosia:2013:phase}
V.~Nicosia, P.~E. V{\'e}rtes, W.~R. Schafer, V.~Latora, and E.~T. Bullmore.
\newblock Phase {Transition} in the {Economically} {Modeled} {Growth} of a
  {Cellular} {Nervous} {System}.
\newblock {\em Proc. Natl. Acad. Sci. USA}, 110(19):7880--7885, May 2013.

\bibitem{pan:2011:path}
R.~K. Pan and J.~Saram{\"a}ki.
\newblock Path {Lengths}, {Correlations}, and {Centrality} in {Temporal}
  {Networks}.
\newblock {\em Phys. Rev. E}, 84(1):016105, July 2011.

\bibitem{roth:2012:long-time}
Camille Roth, Soong~Moon Kang, Michael Batty, and Marc Barthelemy.
\newblock A {Long}-time {Limit} for {World} {Subway} {Networks}.
\newblock {\em Journal of The Royal Society Interface}, page rsif20120259, May
  2012.

\bibitem{scellato:2013:evaluating}
S.~Scellato, I.~Leontiadis, C.~Mascolo, P.~Basu, and M.~Zafer.
\newblock Evaluating {Temporal} {Robustness} of {Mobile} {Networks}.
\newblock {\em IEEE Trans. Mob. Comput.}, 12(1):105--117, January 2013.

\bibitem{schneider:2011:suppressing}
C.~M. Schneider, T.~Mihaljev, S.~Havlin, and H.~J. Herrmann.
\newblock Suppressing {Epidemics} with a {Limited} {Amount} of {Immunization}
  {Units}.
\newblock {\em Phys. Rev. E}, 84(6):061911, December 2011.

\bibitem{sole:2008:robustness}
R.~V. Sol{\'e}, M.~Rosas-Casals, B.~Corominas-Murtra, and S.~Valverde.
\newblock Robustness of the {European} {Power} {Grids} under {Intentional}
  {Attack}.
\newblock {\em Phys. Rev. E}, 77(2):026102, February 2008.

\bibitem{squartini:2013:reciprocity}
T.~Squartini, F.~Picciolo, F.~Ruzzenenti, and D.~Garlaschelli.
\newblock Reciprocity of {Weighted} {Networks}.
\newblock {\em Sci. Rep.}, 3:02729, September 2013.

\bibitem{stam:2014:modern}
C.~J. Stam.
\newblock Modern {Network} {Science} of {Neurological} {Disorders}.
\newblock {\em Nat. Rev. Neurosci.}, 15(10):683--695, 2014.

\bibitem{sur:2015:attack}
S.~Sur, N.~Ganguly, and A.~Mukherjee.
\newblock Attack {Tolerance} of {Correlated} {Time}-varying {Social} {Networks}
  with {Well}-defined {Communities}.
\newblock {\em Physica A}, 420:98--107, February 2015.

\bibitem{tang:2012:stop:}
J.~Tang, H.~Kim, C.~Mascolo, and M.~Musolesi.
\newblock {STOP}: {Socio}-{Temporal} {Opportunistic} {Patching} of {Short}
  {Range} {Mobile} {Malware}.
\newblock In {\em Proc. {IEEE} {International} {Symposium} on a {World} of
  {Wireless}, {Mobile} and {Multimedia} {Networks}}, pages 1--9, 2012.

\bibitem{tang:2009:temporal}
J.~Tang, M.~Musolesi, C.~Mascolo, and V.~Latora.
\newblock Temporal {Distance} {Metrics} for {Social} {Network} {Analysis}.
\newblock In {\em Proc. 2nd {Workshop} on {Online} {Social} {Networks}}, pages
  31--36. ACM, 2009.

\bibitem{tang:2010:small-world}
J.~Tang, S.~Scellato, M.~Musolesi, C.~Mascolo, and V.~Latora.
\newblock Small-world {Behavior} in {Time}-varying {Graphs}.
\newblock {\em Phys. Rev. E}, 81(5):055101, 2010.

\bibitem{thurner:2013:debtrank-transparency:}
S.~Thurner and S.~Poledna.
\newblock {DebtRank}-transparency: {Controlling} {Systemic} risk in {Financial}
  {Networks}.
\newblock {\em Sci. Rep.}, 3:01888, May 2013.

\bibitem{trajanovski:2015:finding}
S.~Trajanovski, F.~A. Kuipers, A.~Ili{\'c}, J.~Crowcroft, and P.~Van Mieghem.
\newblock Finding {Critical} {Regions} and {Region}-{Disjoint} {Paths} in a
  {Network}.
\newblock {\em IEEE/ACM Trans. Netw.}, 23(3):908--921, June 2015.

\bibitem{trajanovski:2013:robustness}
S.~Trajanovski, J.~Mart{\'i}n-Hern{\'a}ndez, W.~Winterbach, and P.~V. Mieghem.
\newblock Robustness {Envelopes} of {Networks}.
\newblock {\em J. Complex Networks}, 1(1):44--62, June 2013.

\bibitem{trajanovski:2012:error}
S.~Trajanovski, S.~Scellato, and I.~Leontiadis.
\newblock Error and {Attack} {Vulnerability} of {Temporal} {Networks}.
\newblock {\em Phys. Rev. E}, 85(6):066105, June 2012.

\bibitem{varier:2011:neural}
S.~Varier and M.~Kaiser.
\newblock Neural {Development} {Features}: {Spatio}-temporal {Development} of
  the {Caenorhabditis} {Elegans} {Neuronal} {Network}.
\newblock {\em PLoS Comput. Biol.}, 7(1):E1001044, January 2011.

\bibitem{varshney:2011:structural}
L.~R. Varshney, B.~L. Chen, E.~Paniagua, D.~H. Hall, and D.~B. Chklovskii.
\newblock Structural {Properties} of the {Caenorhabditis} {Elegans} {Neuronal}
  {Network}.
\newblock {\em PLoS Comput. Biol.}, 7(2):E1001066, February 2011.

\bibitem{wang:2014:studentlife:}
R.~Wang, F.~Chen, Z.~Chen, T.~Li, G.~Harari, S.~Tignor, X.~Zhou, D.~Ben-Zeev,
  and A.~T. Campbell.
\newblock {StudentLife}: {Assessing} {Mental} {Health}, {Academic}
  {Performance} and {Behavioral} {Trends} of {College} {Students} {Using}
  {Smartphones}.
\newblock In {\em Proc. 2014 {Conference} on {Ubiquitous} {Computing}},
  {UbiComp} '14, pages 3--14. ACM, 2014.

\bibitem{white:1986:structure}
J.~G. White, E.~Southgate, J.~N. Thomson, and S.~Brenner.
\newblock The {Structure} of the {Nervous} {System} of the {Nematode}
  {Caenorhabditis} elegans.
\newblock {\em Philos. Trans. R. Soc. London, Ser. B}, 314(1165):1--340,
  November 1986.

\bibitem{williams:2016:data}
M.~J. Williams and M.~Musolesi.
\newblock Data from: {Spatio}-temporal networks: reachability, centrality, and
  robustness.
\newblock {\em Dryad Digital Repository}, 2016.

\bibitem{williams:2016:visualisations}
M.~J. Williams and M.~Musolesi.
\newblock Visualisations from: {Spatio}-temporal networks: reachability,
  centrality, and robustness.
\newblock {\em Figshare Digital Repository}, 2016.

\bibitem{yook:2002:modeling}
S.-H. Yook, H.~Jeong, and A.-L. Barab{\'a}si.
\newblock Modeling the {Internet}'s {Large}-scale {Topology}.
\newblock {\em Proc. Natl. Acad. Sci. USA}, 99(21):13382--13386, October 2002.

\end{thebibliography}


\end{document}